\documentclass[a4paper,twocolumn,superscriptaddress,pre,nofootinbib]{revtex4-1}
\usepackage{xcolor}
\usepackage{hyperref}
\hypersetup{
	colorlinks,
	linkcolor={red!90!black},
	citecolor={black!10!blue},
	urlcolor={blue!80!black}
}

\usepackage{tikz}
\usepackage{amssymb}
\usepackage{amsmath}
\usepackage{graphicx}
\usepackage{multirow}
\usepackage[utf8]{inputenc}

\renewcommand{\r}{\textcolor{red}}
\renewcommand{\d}{\mathrm{d}}
\newcommand{\pmin}{p_\text{min}}
\newcommand{\Tmct}{T_\text{\tiny{MCT}}}

\begin{document}
	\title{The structure-dynamics feedback mechanism governs the glassy dynamics in epithelial monolayers}
	
	\author{Satyam Pandey}
	\affiliation{Tata Institute of Fundamental Research, Hyderabad - 500046, India}
	
	\author{Soumitra Kolya}
	\affiliation{Tata Institute of Fundamental Research, Hyderabad - 500046, India}
	
	\author{Padmashree Devendran}
	\affiliation{Tata Institute of Fundamental Research, Hyderabad - 500046, India}
	
	\author{Souvik Sadhukhan}
	\affiliation{Tata Institute of Fundamental Research, Hyderabad - 500046, India}
	
	\author{Tamal Das}
	\email{tdas@tifrh.res.in}
	\affiliation{Tata Institute of Fundamental Research, Hyderabad - 500046, India}
	
	\author{Saroj Kumar Nandi}
	\email{saroj@tifrh.res.in}
	\affiliation{Tata Institute of Fundamental Research, Hyderabad - 500046, India}


\begin{abstract}  
The glassy dynamics in confluent epithelial monolayers is crucial for several biological processes, such as wound healing, embryogenesis, cancer progression, etc. Several experiments have indicated that, unlike particulate systems, the glassy dynamics in these systems correlates with the static properties and shows a readily-found sub-Arrhenius relaxation. However, whether the statics-dynamics correlation is only qualitative or can provide quantitative predictions and what leads to the sub-Arrhenius relaxation remains unclear. We apply a particular analytical theory of glassy dynamics, the mode-coupling theory (MCT) that predicts dynamics using static properties alone as input, to the confluent systems. We demonstrate the remarkable applicability of MCT in simulations of the Vertex model and experiments on Madin-Darby Canine Kidney cells and show the quantitative nature of the structure-dynamics correlation in these systems. Our results elucidate that the structure-dynamics feedback mechanism of MCT, and not the barrier crossing mechanism, dominates the glassy dynamics in these systems where the relaxation time diverges as a power law with a universal exponent of $3/2$. This slower-than-exponential divergence naturally explains the sub-Arrhenius relaxation dynamics in these systems. The quantitative nature of the structure-dynamics correlation also suggests the possibility of describing various complex biological processes, such as cell division and apoptosis, via the static properties of the systems, such as cell shape or shape variability.
\end{abstract}
\maketitle

\section{Introduction}
 Glassy dynamics, primarily known for the dense aggregates of inanimate particles, is also indispensable for many crucial processes in living organisms \cite{fabry2001,atia2021,park2015a,angelini2011,activereview}. As wounds heal \cite{vishwakarma2020,poujade2007,tetley2018}, embryos develop \cite{spurlin2019,mongera2018}, cancer progresses \cite{friedl2003a,friedl2009b}, or asthma advances \cite{park2015a}, the collective cellular dynamics becomes glassy: the system shows dynamical heterogeneity \cite{angelini2011,garcia2015,vishwakarma2020}, mean-square displacement goes from sub-diffusive at intermediate times to diffusive at longer times \cite{park2015a,malinverno2017,palamidessi2019}, distribution of particle displacement becomes non-Gaussian \cite{giavazzi2017}, etc. The main characteristic of glassiness is the extreme dynamical slowdown, from a liquid-like flowing to a solid-like jammed state without much discernible change in the static properties \cite{berthier2011a,berthier2019,activereview}: a snapshot of a glass and that of a liquid look nearly identical. However, this fundamental feature seems to be different in epithelial systems. 
 Several experiments and simulations have indicated that the glassy dynamics in these systems correlates with the static properties, such as cell shape \cite{park2015a,bi2016,li2021,atia2018,sadhukhan2021a,sadhukhan2022,sadhukhan2024,arora2024}. Furthermore, unlike most particulate systems, cellular monolayer exhibits a remarkable sub-Arrhenius, i.e., slower than exponential relaxation \cite{sussman2018b,sadhukhan2021a,li2021}. Is there a qualitative difference between the glassy dynamics in these systems and particulate systems? What is the mechanism leading to this sub-Arrhenius relaxation? Is the statics-dynamics correlation merely qualitative, or does it have more profound significance?

Epithelial cellular monolayers have a distinctive characteristic compared to particulate systems. They are confluent, i.e., the cells fill the entire space, and the packing fraction remains unity where the glassy dynamics is controlled by intercellular interaction and various active forces.
The theoretical models represent cells as polygons and can be either discrete lattice-based, such as the cellular Potts model \cite{graner1992,hogeweg2000}, or continuum, such as the Vertex or Voronoi models \cite{honda1980,farhadifar2007,fletcher2014,bi2016}. The energy function governing the system properties \cite{honda1980,albert2016} is
\begin{equation}\label{energyfunction}
	\mathcal{H}=\sum_{i=1}^N [\lambda_A(a_i-a_0)^2+\lambda_P(p_i-p_0)^2],
\end{equation}
where $N$ is the total number of cells, $\lambda_A$ and $\lambda_P$ are the area and perimeter moduli, $a_0$ and $p_0$ are the target area and target perimeter, and $a_i$ and $p_i$ are the individual cell area and perimeter. We set $a_0=1$ and use $\sqrt{a_0}$ as the unit of length. Experiments show that the height of a cell monolayer remains nearly the same \cite{farhadifar2007}, and the cell cytoplasm behaves like an incompressible fluid \cite{jacques2015}. These two aspects lead to the area constraint in Eq. (\ref{energyfunction}). In addition, the properties of intercellular interaction mediated via different junctional proteins and that of a thin layer of cytoplasm, the cell cortex, lead to the perimeter term. We also need a temperature $T$ that comes from various active processes. However, within the models, $T$ is an equilibrium temperature. Such a description agrees remarkably well with experiments \cite{park2015a,bi2014,atia2018,sadhukhan2022}. There are several confluent models, and they vary in their implementation details. However, they are similar from the perspective of glassy dynamics \cite{bi2014,park2015a,bi2016,sussman2018a,sadhukhan2021a}. In this work, we focus on the Vertex model for our simulations.

The physics of glassy dynamics generally refers to barrier crossing between various metastable states. In an Angell plot representation \cite{angell1991,angell1995}, plotting relaxation time as a function of $T_g/T$, where $T_g$ is the glass transition temperature, a straight line refers to Arrhenius relaxation. Curves below this line signify super-Arrhenius and above it sub-Arrhenius relaxation [as marked in Fig. \ref{mct_sim_comp}(f)]. Simulation results have shown that the distinctive potential arising from the perimeter constraint in Eq. (\ref{energyfunction}) plays a crucial role in the dynamics of epithelial systems. The geometric constraint of space-filling polygons leads to a minimum possible perimeter, $\pmin$. 
When $p_0 \lesssim\pmin$, the cell boundaries are primarily straight, and cells cannot satisfy the perimeter constraint in Eq. (\ref{energyfunction}), the system shows sub-Arrhenius relaxation \cite{sussman2018b,sadhukhan2021a,li2021}.
The relaxation time in the sub-Arrhenius regime grows {\it slower} than exponential, implying a different mechanism than barrier crossing for the glassiness in this regime. But what is that mechanism? How does it affect the relaxation dynamics? What are the implications of the structure-dynamics correlation?

In this work, we first show that a specific analytical theory of glassy dynamics, the mode-coupling theory (MCT) \cite{gotzebook,das2004,gotze1992,reichman2005,geszti1983}, works remarkably well for confluent systems in the sub-Arrhenius regime. MCT assumes that the static properties of the system are known, and taking the static properties as input, it predicts the dynamics. MCT posits a feedback mechanism, where the structure affects the dynamics that again influences the structure for the glassy dynamics. This feedback mechanism leads to power law divergence of the relaxation time and naturally explains the sub-Arrhenius relaxation. We demonstrate this quantitative nature of the structure-dynamics correlation in these systems via MCT for both simulations on the Vertex model and experiments on Madin-Darby Canine Kidney (MDCK) cell monolayers. 
Our results imply that various biological functions, such as cell division and apoptosis that can affect the dynamics can be described in terms of the static properties of the system.

\begin{figure}
	\centering
	\includegraphics[width=8.6cm]{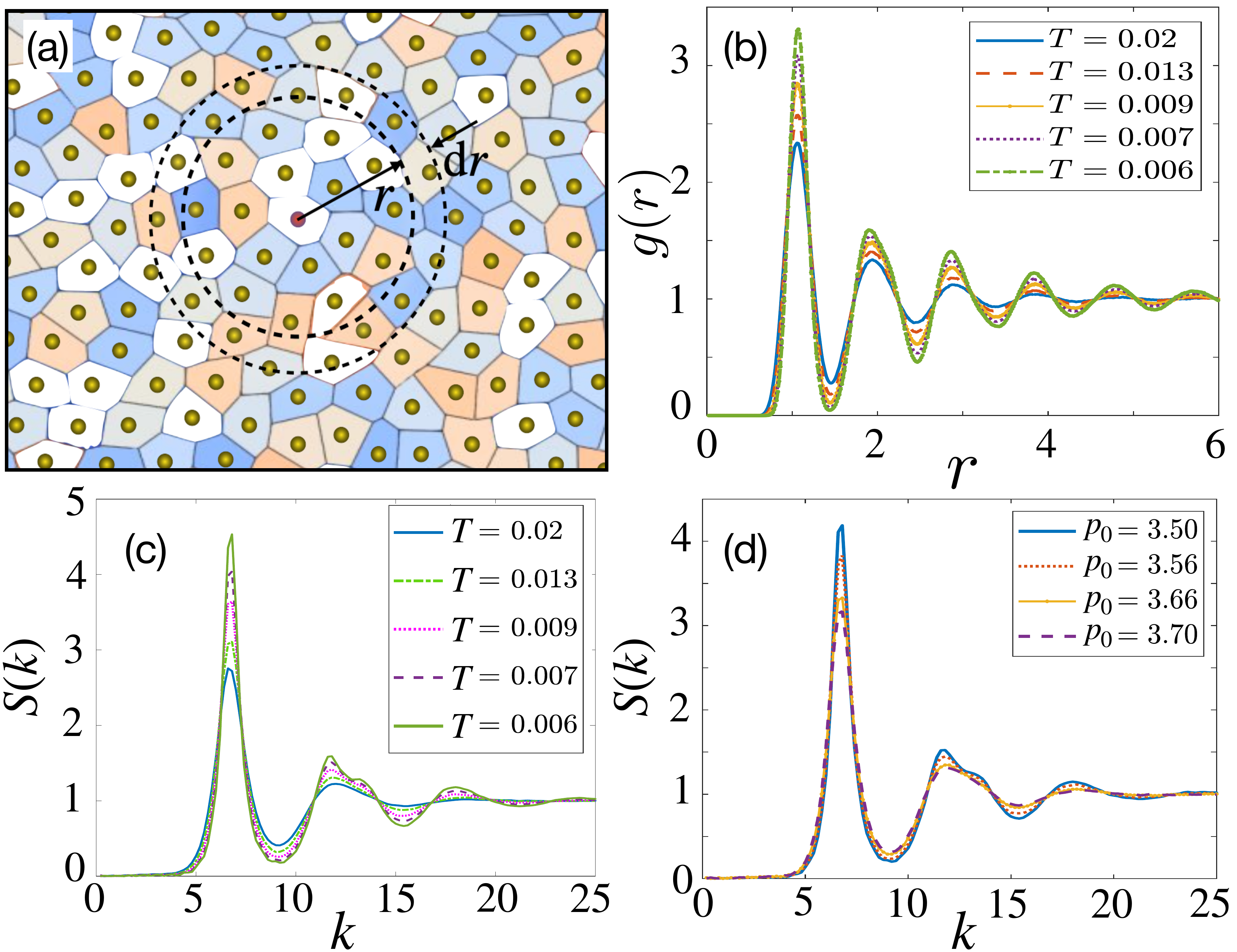}
	\caption{Static properties of confluent systems. (a) Schematic presentation of the calculation of $g(r)$. We represent the position of the cells by their centers of mass at $\mathbf{r}^{\text{cm}}_i$. Starting from a particular cell center as origin, $g(r)$ gives the probability distribution of finding another cell in the annular region of radii $r$ and $r+\d r$. (b) We show the radial distribution function, $g(r)$, as a function of $r$ at various $T$ and $p_0=3.60$. (c) We obtain $S(k)$ by taking the Fourier transform of $g(r)$ according to Eq. (\ref{FT_Sk_gr}). $S(k)$ peak height increases as $T$ decreases. (d) $S(k)$ at a constant $T=0.009$ shows increase in peak height as we decrease $p_0$.}
	\label{gr_sk}
\end{figure}

\section*{Results}

We first show in our simulations on the Vertex model that MCT applies quite well in the sub-Arrhenius regime and 
provides the dynamics using the static properties as input: we compute the static structure factor, give it as input to MCT, obtain the dynamics from the theory, and compare it with that in simulations. We then demonstrate the applicability of MCT in our experiments on MDCK cells.

\subsection*{Statics and dynamics in simulation and comparison with MCT}
We simulate the Vertex model for various values of $p_0$ and $T$ (see Sec. \ref{simulationdetails} for details). MCT requires the static information of cell positions in the form of the radial distribution function, $g(r)$, defined in Sec. \ref{definitions}. We designate each cell by its center of mass, $\mathbf{r}^{\text{cm}}_i$, where $i=1,N$ (see supplementary material (SM), Sec. SII for details). Using the center of mass for the cells, we obtain $g(r)$ as schematically shown in Fig. \ref{gr_sk}(a): we start from a particular cell center as the origin, count the number of cells within the annular region of radii $r$ and $r+\d r$, and divide it by the average number of cells within this area. We repeat this process for all cells and then take the average. $g(r)$ gives the probability density of finding a cell at a distance $r$ starting from another cell as the origin. We Fourier transform $g(r)$ as in Eq. (\ref{FT_Sk_gr}), and obtain the static structure factor, $S(k)$. Figure \ref{gr_sk}(b) shows several $g(r)$ for various $T$ at $p_0=3.6$; the height of the first peak increases as $T$ decreases. Figure \ref{gr_sk}(c) shows the corresponding $S(k)$ for the same $g(r)$ as in (b); $S(k)$ also has an increasing peak height with decreasing $T$. Figure \ref{gr_sk}(d) shows $S(k)$ at a fixed $T$ but different values of $p_0$: as $p_0$ decreases, $S(k)$ peak height increases.

\begin{figure}[t!]
	\includegraphics[width=8.6cm]{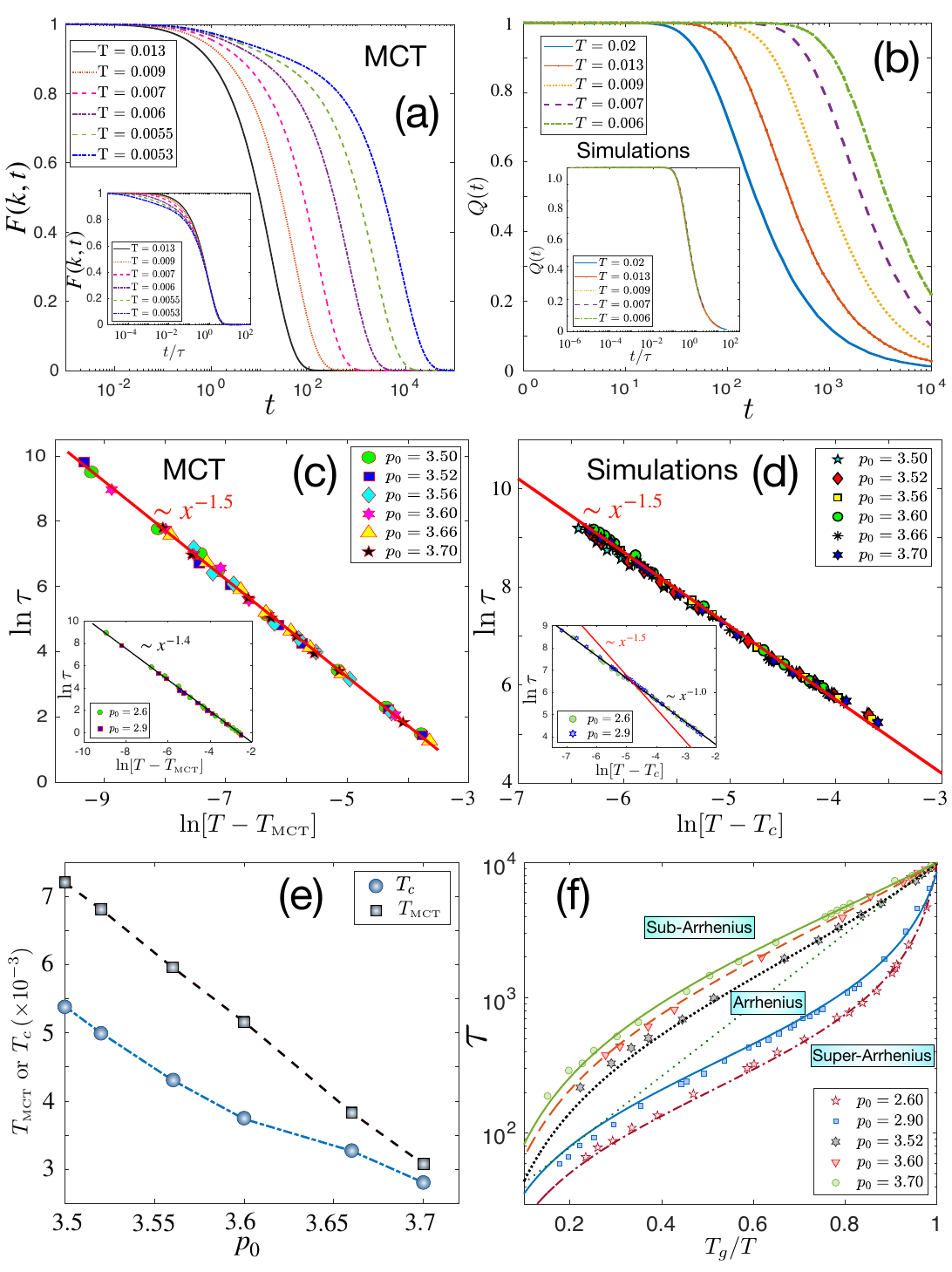}
	\caption{Comparison of MCT and simulation results. (a) The intermediate scattering function, $F(k,t)$, obtained via MCT using the simulation $S(k)$ as input. {\bf Inset:} The time-temperature superposition prediction of MCT: $F(k,t)$ as a function of $t/\tau(T)$ provides data collapse to a master curve in the $\alpha$-regime. (b) $Q(t)$ from the vertex model simulation as a function of time $t$ for several $T$. {\bf Inset:} Simulation data agree with the time-temperature superposition prediction of MCT. (c) MCT predicts an algebraic divergence of $\tau$ at $T_\text{\tiny MCT}$ with universal exponent, Eq. (\ref{MCTprediction}). The plot of $\ln\tau$ as a function of $\ln[T-T_\text{\tiny MCT}]$ shows that data for various $p_0$ follow the same master curve demonstrating the predictions. {\bf Inset:} MCT data of $\tau$ at lower values of $p_0$ show a slightly different $\gamma$. (d) Fitting the simulation data of $\tau$ with the MCT power-law form gives the transition temperature $T_c$ and the other parameters. The plot of $\ln\tau$ as a function of $\ln[T-T_c]$ again shows a master curve confirming the MCT prediction of universal $\gamma$. The lines show that $\gamma$ in MCT and simulation data are the same. {\bf Inset:} Simulation data of $\tau$ at lower $p_0$ yield $\gamma=1.0$ that is different from the MCT result for similar $p_0$. (e) MCT over-estimates the transition, $T_\text{\tiny MCT}$ is higher compared to $T_c$. The difference decreases as $p_0$ increases. (f) Angell plot of $\tau$ as a function of $T_g/T$ where we have defined $T_g$ via $\tau(T_g)=10^4$. Symbols are simulation data, and the lines are the corresponding MCT plots.}
	\label{mct_sim_comp}
\end{figure}

We next use the simulation data of $S(k)$ and the corresponding $T$ as input to MCT and numerically solve the theory as an initial value problem to obtain the intermediate scattering function, $F(k,t)$ (defined in Eq. \ref{fofkt}), where $k$ is the wave vector, as a function of time $t$. Within MCT, the principal effects of varying parameters enter via $S(k)$. The bare $T$ appearing in the MCT equations (see SM, Sec. SIII) gives the time scale. Figure \ref{mct_sim_comp}(a) shows the MCT results for $F(k,t)$ at $k=6.8$, corresponding to the first peak of $S(k)$. In simulations, computation of $F(k,t)$ is costly as it requires large averaging due to fluctuation \cite{pareek2023}. We can obtain similar information about the dynamics via a related observable, the overlap function, $Q(t)$, Eq. (\ref{Qoftdef}). We show the behavior of $Q(t)$ in simulations in Fig. \ref{mct_sim_comp}(b). The decay of $F(k,t)$ and $Q(t)$ gets slower as $T$ decreases. From the decay of the auto-correlation functions, we can define a relaxation time, $\tau$, via $Q(\tau)=0.3$ and $F(k,\tau)=0.3$. $\tau$ increases as the system becomes more glassy when $T$ decreases. We define the glass transition temperature $T_g$ when $\tau$ becomes $10^4$, i.e., $\tau(T_g)=10^4$. One characteristic of glassy dynamics is a slow decay of the auto-correlation function at long times, known as the $\alpha$-regime. MCT predicts a time-temperature superposition principle in the $\alpha$-regime, i.e., the two-point functions at different $T$ as a function of $t/\tau(T)$ collapse to a master curve in the $\alpha$-regime. We have demonstrated this property in the insets of Fig. \ref{mct_sim_comp}(a) for $F(k,t)$ and Fig. \ref{mct_sim_comp}(b) for $Q(t)$.

We now look at the most rigorous test of MCT. The primary prediction of MCT is a power-law divergence of $\tau$ with a universal exponent \cite{gotzebook,gotze1992,berthier2011a,janssen2018}. 
As Fig. \ref{mct_sim_comp}(a) shows, the decay of $F(k,t)$ becomes progressively slower as we lower $T$. Eventually, below a particular $T$, known as $T_\text{\tiny{MCT}}$, $F(k,t)$ remains stuck at a finite value: this is the non-ergodicity transition of MCT (SM Fig. S2). The non-ergodicity transition is a genuine phase transition within the theory that describes glass transition as a critical phenomenon with $T_\text{\tiny MCT}$ as the critical point. Just like theories of critical phenomena \cite{kardarbook}, MCT also predicts power-law divergence of $\tau$ as $T$ approaches $T_\text{\tiny{MCT}}$:
\begin{equation}\label{MCTprediction}
\tau = \frac{A}{(T - T_\text{\tiny{MCT}})^\gamma } ,
\end{equation}
where $A$ is a constant and $\gamma$ is a universal exponent. It is well-known that the non-ergodicity transition itself is spurious as some other mechanism takes over for $T<T_\text{\tiny{MCT}}$.
However, the theory works in the range of $T$ where most simulations and experiments operate. In the regime of its validity, if we fit simulation and experimental data with the MCT prediction, Eq. (\ref{MCTprediction}), we obtain a slightly lower $T$ than $\Tmct$ as the critical point; let us call it $T_c$.

We first focus on the range of $p_0$ between $3.5$ and $3.7$, corresponding to the sub-Arrhenius regime \cite{sadhukhan2021a,sussman2018b,li2021}. For a particular $p_0$, we get $\tau$ for several values of $T$ and then vary $p_0$. We fit the data of $\tau$ from both MCT and simulation with Eq. (\ref{MCTprediction}) to obtain $\Tmct$ (or $T_c$), $A$, and $\gamma$. We then plot the data for $\tau$ as a function of $(T-\Tmct)$ for the MCT results and as a function of $(T-T_c)$ for the simulation data. The power-law prediction of MCT, Eq. (\ref{MCTprediction}), implies that the data in log-log representation will be linear: 
\begin{equation}\label{powerlawform}
\ln\tau = \ln A - \gamma \ln(T-T_\text{\tiny{MCT}}).
\end{equation}
Figures \ref{mct_sim_comp}(c) and (d) show that the data collapse to a master curve for both MCT and simulation. This data collapse implies that $A$ and $\gamma$ are the same for different $p_0$ in this range. Furthermore, $\gamma=3/2$ is the same for both MCT and simulation [represented by the solid lines in Figs. (\ref{mct_sim_comp}c) and (\ref{mct_sim_comp}d)]. Thus, the MCT predictions - power-law behavior of $\tau$ with universal $\gamma$ - agree remarkably well with the simulations. In contrast, although the power-law prediction of MCT holds in particulate systems, $\gamma$ varies for different systems \cite{gotzebook,das2004}.

We have also investigated the lower range of $p_0$, where the system shows super-Arrhenius behavior \cite{sadhukhan2021a,sadhukhan2024}: the insets of Figs. \ref{mct_sim_comp} (c) and (d) show that $\gamma$ becomes different in this regime, and the agreement between MCT and simulation becomes poor. For example, for $p_0=2.6$ and $2.9$, the MCT data suggest $\gamma\simeq1.4$, whereas the simulation data show $\gamma \simeq 1$. This scenario is similar to the usual particulate systems where the exponent usually vary. We know that the RFOT theory applies in this regime \cite{sadhukhan2021a,sadhukhan2024}, and activated barrier crossing dominates the dynamics. $\gamma$ becomes strongly $p_0$-dependent for even lower values of $p_0$. We also find that both $\Tmct$ and $T_c$ decrease nearly linearly with an increase in $p_0$ (Fig. \ref{mct_sim_comp}e). Since $p_0$ parameterizes intercellular interaction, this trend of $\Tmct$ or $T_c$ represents the effects of changing interaction potential.

MCT posits a structure-dynamics feedback mechanism for the glassy slowing down: structure influences dynamics that feeds back to the structure and leads to the diverging $\tau$ \cite{geszti1983,gotze1992,gotzebook,das2004}. The remarkable agreement of the dynamics with MCT implies that this feedback mechanism, and not the barrier-crossing scenario, controls the glassy character in these systems. This mechanism leads to a power-law variation of $\tau$; thus, it naturally leads to sub-Arrhenius, i.e., slower than exponential relaxation dynamics. Figure \ref{mct_sim_comp}(f) shows the MCT and the corresponding simulation data in the Angell plot representation. Our results elucidate that the sub-Arrhenius relaxation dynamics and structure-dynamics correlation are congruous.


\begin{figure*}[t!]
	\centering
	\includegraphics[width=0.9\textwidth]{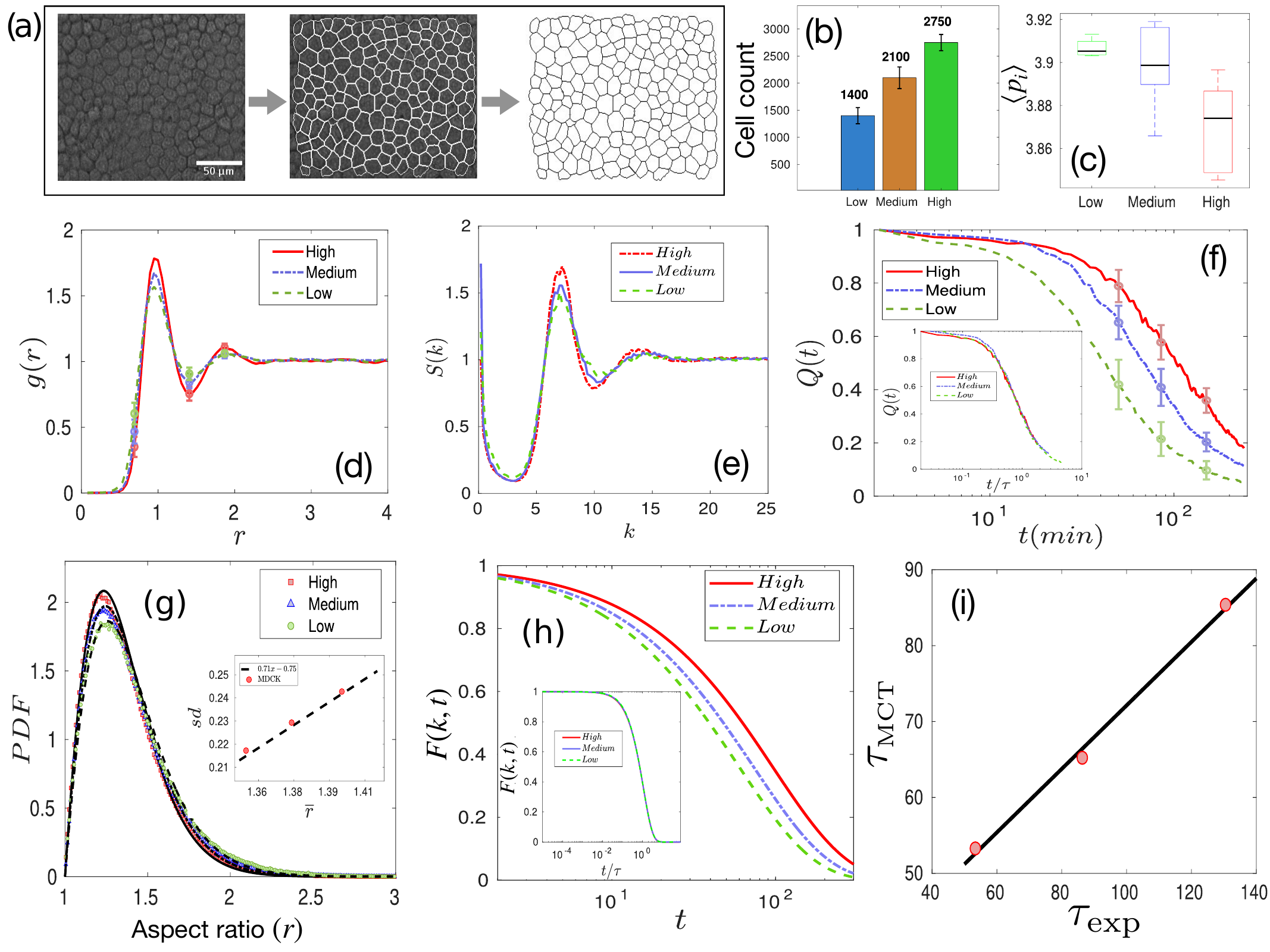}
	\caption{Structure-dynamics correlation in experiments. (a) An illustration of the segmentation and data analysis, we give the raw image to the cellpose software for segmentation and analyze the segmented images. (b) Number of cells in the field of view for the three densities. Error bars show the number fluctuation. (c) The observed shape index, $\langle p_i\rangle$, for the three systems are such that they are in the fluid regime. (d) The first peak of the radial distribution function, $g(r)$, increases as cell number density grows. (e) The corresponding static structure factor, $S(k)$, also shows a similar trend as that of $g(r)$. (f) The decay of the overlap function, $Q(t)$, becomes slower with increasing density. {\bf Inset}: $Q(t)$ shows data collapse when we plot them as a function of $t/\tau$. (g) We fit the experimental data of cell shape variability with the CSD function, Eq. (\ref{csdfunc}) and obtain the values of $\alpha$ that vary inversely with $T$. {\bf Inset}: $sd$ vs $\bar{r}$ follows a universal relation (see text). (h) We use $S(k)$ as input to MCT and obtain the dynamics via $F(k,t)$. {\bf Inset}: Time-temperature superposition principle similar to that in the experiment (Fig. f). (i) $\tau$ from MCT, $\tau_\text{MCT}$, is linearly proportional to $\tau_\text{exp}$ obtained in experiment.}
	\label{experimental_comp}
\end{figure*}

\subsection*{Applicability of MCT in experiments}
We now show that MCT can also predict the dynamics from the static structure in experiments. For this purpose, we conducted various cell culture experiments using MDCK epithelial cells. Changing $T$ is not convenient in cellular experiments. We, therefore, measured the dynamics by varying cellular number density, $\rho$, that changes the dynamics \cite{park2015a,garcia2015}. We emphasize that $\rho$ does not directly affect the dynamics in these systems; it changes the inter-cellular adhesion that controls the dynamics \cite{garcia2015}. For better averaging, we wanted various systems with constant $\rho$. However, precise control of $\rho$ is challenging as it sensitively depends on the number of seeding cells. We start with seeding a small number of cells so that the system becomes confluent over time. 
However, if the number of seeding cells is too small, the system does not attain confluency within the experimental time scale. By contrast, if it is too large, the system becomes overconfluent, and cells start extruding out. To avoid these difficulties, we chose three densities such that they give distinct dynamical properties. For data acquisition, we take images at intervals of 2.5 minutes for 4 hours. We used the Cellpose software \cite{stringer2021} for segmentation and data analysis (see the SM for more details). Figure \ref{experimental_comp}(a) shows a typical snapshot of the monolayer and the image analysis.

For the three densities, Fig. \ref{experimental_comp}(b) shows the average number of cells in the field of view, as well as the variance in the cell number (see SM for details). Figure \ref{experimental_comp}(c) shows the average observed perimeter, $\langle p_i/\sqrt{a_i} \rangle$, for the three systems: the values are such that the systems are in the fluid regime \cite{park2015a,bi2016,sadhukhan2024}. Following the same method as in the simulations, we have represented cells by their centers of mass, and calculated $g(r)$. We find that the height of the first peak of $g(r)$ increases with increasing $\rho$ (Fig. \ref{experimental_comp}d). We take a Fourier transform of $g(r)$ and calculate $S(k)$ via Eq. (\ref{FT_Sk_gr}). Figure \ref{experimental_comp}(e) shows $S(k)$ for the systems, similar to $g(r)$ data, the height of the first peak of $S(k)$ also increases as $\rho$ increases. We highlight one unusual property compared to particulate systems, $S(k)$ increases as $k\to0$; this comes from the fluctuation induced by cell division and apoptosis \cite{giavazzi2017,giavazzi2018}. 
We can use this $S(k)$ as input to MCT to obtain dynamics. We also measure the relaxation dynamics in experiments via $Q(t)$. Consistent with existing results \cite{park2015a,garcia2015}, the decay of $Q(t)$ becomes slower, and $\tau$ becomes higher, as $\rho$ increases (Fig. \ref{experimental_comp}f).

To compute dynamics via MCT, we still need the parameter $T$. For this, we apply the cell shape variability theory, considering an effective equilibrium scenario.
We first calculate the gyration tensor for each cell; diagonalization of this tensor gives two eigenvalues. The square root of the ratio of the two eigenvalues gives the aspect ratio, $r$. Reference \cite{sadhukhan2022} has derived the cell-shape distribution (CSD) function, $P(r)$, that depends on a single parameter $\alpha$:
\begin{equation} \label{csdfunc}
P(r)=\frac{1}{\mathcal{N}}\left(r+\frac{1}{r}\right)^{3 / 2}\left(1-\frac{1}{r^2}\right) e^{-\alpha\left(r+\frac{1}{r}\right)}
\end{equation}
where $\alpha \propto \lambda_P\left(1 - K p_0\right) / T$, $K$ is a constant, and $\mathcal{N}$ is the normalization constant. Thus, $\alpha$ is inversely proportional to $T$. We fit the CSD function with the experimental data and obtain the values of $\alpha$ (Fig. \ref{experimental_comp}g). One consequence of the CSD function is that the average cell shape, $\bar{r}$, and the variance, $sd$, follow a universal relation: $sd=0.71 \bar{r}-0.75$. The excellent agreement between theory and experiment for this universal prediction [inset of Fig. \ref{experimental_comp}(g)] shows the applicability of the theory. Therefore, we take $T$ as the inverse of $\alpha$, i.e., $T=C/\alpha$, where $C$ is a constant. We determine $C$ using the experimental data. The main effect of changing $T$ or density enters the MCT via $S(k)$. The quantitative value of $T$ only determines the time scale and does not affect the functional dependence of $\tau$ on $T$. To determine the value of $C$, we solve MCT with varying $C$ and using the static parameters corresponding to the lowest density system. We chose the value of $C\simeq 0.0044$ such that $\tau$ from MCT and the experiment match at this density. We have checked that the qualitative results do not change for other values of $C$. We then use $S(k)$ and $T$ for the experimental systems and solve the MCT to obtain the dynamics.  We show the $F(k,t)$ for a particular $k$, corresponding to the $S(k)$ maximum, for the three different systems in Fig. \ref{experimental_comp}(h). 

We show the time-temperature superposition, i.e., the data collapse of the auto-correlation functions when we plot them as functions of $t/\tau$, for the experiments (inset, Fig. \ref{experimental_comp}f) and MCT (inset, Fig. \ref{experimental_comp}h). We have also checked that the experimental data are consistent with the power-law form, Eq. (\ref{powerlawform}), with exponent 3/2 (See SM Fig. S6). For a quantitative comparison of the experimental $\tau$ with the MCT results, we plot the two relaxation times against each other. The straight line signifies that they are proportional to each other, implying that MCT does predict the dynamics, taking the static property as an input. Thus, MCT works remarkably well for these systems, and the feedback mechanism controls the glassy dynamics leading to the sub-Arrhenius relaxation. As we discuss below, the results also have several crucial implications.


\section*{Discussion and conclusion}
We have demonstrated in simulations and experiments of confluent systems that MCT applies remarkably well in their sub-Arrhenius regime. Since MCT takes the static structure alone as input and predicts the dynamics, the applicability of the theory implies the quantitative nature of the structure-dynamics correlation indicated by several past works \cite{park2015a,atia2018,sadhukhan2022}.  The results presented in this work have several consequences. 
First, the remarkable agreement of MCT with the dynamics, both in simulations and experiments, demonstrates that we can use the power law predictions of MCT to analyze the glassy dynamics in these systems. 
MCT describes the glassy dynamics via a feedback mechanism: a minute change in the static property influences the dynamics that again affects the statics \cite{geszti1983,gotzebook,das2004}. This feedback mechanism leads to a power-law divergence of $\tau$ with a universal exponent, $\gamma$. For the confluent systems in their sub-Arrhenius regime, we find this exponent $\gamma\simeq 3/2$. Our experimental results are consistent with this power-law behavior of $\tau$ and the value of $\gamma$.
 
Second, the applicability of MCT provides straightforward explanations for the nontrivial glassy dynamics of these systems. The feedback mechanism of MCT predicts a power-law divergence of $\tau$. A power law is slower than exponential; therefore, it naturally leads to the sub-Arrhenius relaxation dynamics (Fig. \ref{mct_sim_comp}f). Thus, the structure-dynamics correlation and the sub-Arrhenius relaxation dynamics are synonymous.

Third, it raises the possibility of using a static parameter, such as average cell shape or its variability, to describe cellular functions, such as cell division or apoptosis that affect dynamics. 
We have excluded these processes in our simulations; however, they are present in the experiment. It is well-known that these processes cut off the time scale and fluidize a cellular monolayer \cite{ranft2010}. MCT predicts the relaxation time scale only via the use of static properties. Thus, the effect of these processes must be present within the static description of the system. This expectation is consistent with several past works that have confirmed a strong correlation between cell shape and functions \cite{minc2011,chen1997,jaiswal2006}. We are currently exploring how far this correlation exists and if we can use it for quantitative predictions and develop a theoretical description for cell division and apoptosis using the theory of cell shape variability.

We believe our results will also be interesting to scientists from the field of glassy dynamics in general. MCT is an elegant first-principle theory of glassy dynamics \cite{gotzebook,das2004,reichman2005}. However, it enjoys a mixed fate. It is well-known that MCT fails at low $T$, where it predicts a spurious non-ergodicity transition \cite{gotzebook,cates2006,janssen2018}. Despite this failure, MCT remains popular for several reasons: it is an analytical theory for an immensely complex problem, many colloidal experiments and simulations operate in a regime where MCT is valid, it only requires $S(k)$ as input and therefore applicable for novel systems whose detailed properties are not yet known, etc. It is intriguing that the theory works so well for the sub-Arrhenius regime of confluent systems and might provide crucial insights about the theory itself.
Why MCT works so well for the confluent systems \cite{ruscher2021} remains unclear. One possibility is that the constraint of confluency imposes long-range interaction. Since MCT is a critical theory, this long-range nature of the interaction might favor MCT.

In conclusion, we have shown via the application of MCT that the structure dynamics correlation of epithelial monolayers is quantitative. This implies that the feedback mechanism of MCT governs the glassy dynamics. This mechanism leads to a power law growth of $\tau$ with a universal exponent $\gamma\simeq 3/2$. A power law is slower than exponential: the curve will fall above the Arrhenius line in the Angell plot. Thus, MCT provides a natural explanation of the sub-Arrhenius relaxation dynamics of these systems. We have also demonstrated that MCT works well for our experimental system of the MDCK monolayer. The applicability of MCT to these systems hints at a fascinating aspect of describing cellular processes, such as cell division and apoptosis, that affect dynamics via static observables, such as cell shape.

\section{Materials and Methods}

\subsection{Definitions}
\label{definitions}
Here we provide the definitions of the different variables used in this work. 
The definition of the static structure factor, $S(k)$, at wavevector $k$ is
\begin{equation}
	{S}(k)=\frac{1}{N}\left\langle\sum_{j,l=1}^N e^{-i \mathbf{k} \cdot\left(\mathbf{r}_{{j}}-\mathbf{r}_{{l}}\right)}\right\rangle,
\end{equation}
where $j$ and $l$ are particle indices, and $k=|\mathbf{k}|$. However, the numerical evaluation of $S(k)$ through the above equation is challenging due to fluctuations and requires a lot of averaging. Instead, we first compute direct correlation function $g(r)=\langle \sum_{i\neq0} \delta(|\mathbf{r}-\mathbf(r)_i|)\rangle/\rho$, where $\rho$ is the particle density. We then obtain  $S(k)$ via the Fourier transform as
\begin{equation}\label{FT_Sk_gr}
	{S}(k)=1+\rho \int e^{-i \mathbf{k} \cdot \mathbf{r}}[g(r)-1] d \mathbf{r}+\mathrm{N} \delta_{\mathrm{k}, 0},
\end{equation}
where $\delta_{\mathrm{k}, 0}$ is the Kronecker $\delta$-function. This method yields much smoother data.

In the simulation, we characterize dynamics via the overlap function, $Q(t)$, defined as
\begin{equation}\label{Qoftdef}
	Q(t)=\left\langle\frac{1}{N} \sum_{i=1}^N W\left(a-\left|\mathbf{r}_i(t)-\mathbf{r}_i(0)\right|\right)\right\rangle,
\end{equation}
where $W(x)$ is the Heaviside step function:
\begin{equation}
	W(x)=\begin{cases}
		1 \,\,\, &\text{if }\,\, x>0,\\
		0\,\,\,\ &\text{otherwise}.
	\end{cases}
\end{equation}
The parameter $a$ is set to a constant value of 1.867 throughout the simulation.

Within MCT, we characterize dynamics via the intermediate scattering function, $F(k,t)$, at time $t$,
\begin{equation}\label{fofkt}
	F(k,t)=\frac{1}{N}\langle \rho_k(t)\rho_{-k}(0)\rangle=\frac{1}{N}\sum_{jl}e^{i\mathbf{k}\cdot(\mathbf{r}_j(t)-\mathbf{r}_l(0))},
\end{equation}
where $\rho_k(t)$ is the Fourier mode of density at time $t$.

\subsection{Simulation Details}
\label{simulationdetails}
We conducted the simulations both via a code written by ourselves and by modifying an open-source software, RheoVM \cite{tong2022}. 
We have used a two-dimensional system of size $L\times L$ with $L=30$ and a total number of cells $N=400$ with preferred area $a_0= \frac{L^2}{N}$. We start with a disordered configuration and use Brownian dynamics for the simulation.
The forces resulting from cell shape is $\mathbf{F}_i=-\nabla{\mathbf{r}_i} \mathcal{H}$, with the equation of motion of the $i$th vertex is
\begin{equation}
	\gamma \dot{\mathbf{r}}_i=\mathbf{F}_i + \sqrt{2 D_T} \zeta,
\end{equation}
where $\mathbf{r}_i$ is the position of $i$th vertex, $ D_T=T/\gamma$ is translational diffusivity at $T$, $\gamma$ is the substrate friction. $\zeta$ is a random noise drawn from a normal distribution with zero mean and unit variance. In our simulation, we set the ratio $\lambda_P/\left(\lambda_A a_0\right) \approx 0.044$ and $\gamma= 0.1$. We express length in units of $\sqrt{a_0}$; this allows us to compare various simulation results.

\section{Acknowledgements}

We thank Kabir Ramola and J. J. Fredberg  for discussions. We acknowledge the support of the Department of Atomic Energy, Government of India, under Project Identification No. RTI 4007. SKN thanks SERB for grant via SRG/2021/002014.

\bibliography{activeglassref.bib}

\begin{thebibliography}{54}%
\makeatletter
\providecommand \@ifxundefined [1]{%
 \@ifx{#1\undefined}
}%
\providecommand \@ifnum [1]{%
 \ifnum #1\expandafter \@firstoftwo
 \else \expandafter \@secondoftwo
 \fi
}%
\providecommand \@ifx [1]{%
 \ifx #1\expandafter \@firstoftwo
 \else \expandafter \@secondoftwo
 \fi
}%
\providecommand \natexlab [1]{#1}%
\providecommand \enquote  [1]{``#1''}%
\providecommand \bibnamefont  [1]{#1}%
\providecommand \bibfnamefont [1]{#1}%
\providecommand \citenamefont [1]{#1}%
\providecommand \href@noop [0]{\@secondoftwo}%
\providecommand \href [0]{\begingroup \@sanitize@url \@href}%
\providecommand \@href[1]{\@@startlink{#1}\@@href}%
\providecommand \@@href[1]{\endgroup#1\@@endlink}%
\providecommand \@sanitize@url [0]{\catcode `\\12\catcode `\$12\catcode
  `\&12\catcode `\#12\catcode `\^12\catcode `\_12\catcode `\%12\relax}%
\providecommand \@@startlink[1]{}%
\providecommand \@@endlink[0]{}%
\providecommand \url  [0]{\begingroup\@sanitize@url \@url }%
\providecommand \@url [1]{\endgroup\@href {#1}{\urlprefix }}%
\providecommand \urlprefix  [0]{URL }%
\providecommand \Eprint [0]{\href }%
\providecommand \doibase [0]{http://dx.doi.org/}%
\providecommand \selectlanguage [0]{\@gobble}%
\providecommand \bibinfo  [0]{\@secondoftwo}%
\providecommand \bibfield  [0]{\@secondoftwo}%
\providecommand \translation [1]{[#1]}%
\providecommand \BibitemOpen [0]{}%
\providecommand \bibitemStop [0]{}%
\providecommand \bibitemNoStop [0]{.\EOS\space}%
\providecommand \EOS [0]{\spacefactor3000\relax}%
\providecommand \BibitemShut  [1]{\csname bibitem#1\endcsname}%
\let\auto@bib@innerbib\@empty
\bibitem [{\citenamefont {Fabry}\ \emph {et~al.}(2001)\citenamefont {Fabry},
  \citenamefont {Maksym}, \citenamefont {Butler}, \citenamefont {Glogauer},
  \citenamefont {Navajas},\ and\ \citenamefont {Fredberg}}]{fabry2001}%
  \BibitemOpen
  \bibfield  {author} {\bibinfo {author} {\bibfnamefont {B.}~\bibnamefont
  {Fabry}}, \bibinfo {author} {\bibfnamefont {G.~N.}\ \bibnamefont {Maksym}},
  \bibinfo {author} {\bibfnamefont {J.~P.}\ \bibnamefont {Butler}}, \bibinfo
  {author} {\bibfnamefont {M.}~\bibnamefont {Glogauer}}, \bibinfo {author}
  {\bibfnamefont {D.}~\bibnamefont {Navajas}}, \ and\ \bibinfo {author}
  {\bibfnamefont {J.~J.}\ \bibnamefont {Fredberg}},\ }\href {\doibase
  10.1103/PhysRevLett.87.148102} {\bibfield  {journal} {\bibinfo  {journal}
  {Phys. Rev. Lett.}\ }\textbf {\bibinfo {volume} {87}},\ \bibinfo {pages}
  {148102} (\bibinfo {year} {2001})}\BibitemShut {NoStop}%
\bibitem [{\citenamefont {Atia}\ \emph {et~al.}(2021)\citenamefont {Atia},
  \citenamefont {Fredberg}, \citenamefont {Gov},\ and\ \citenamefont
  {Pegoraro}}]{atia2021}%
  \BibitemOpen
  \bibfield  {author} {\bibinfo {author} {\bibfnamefont {L.}~\bibnamefont
  {Atia}}, \bibinfo {author} {\bibfnamefont {J.~J.}\ \bibnamefont {Fredberg}},
  \bibinfo {author} {\bibfnamefont {N.~S.}\ \bibnamefont {Gov}}, \ and\
  \bibinfo {author} {\bibfnamefont {A.~F.}\ \bibnamefont {Pegoraro}},\ }\href
  {\doibase https://doi.org/10.1016/j.cdev.2021.203727} {\bibfield  {journal}
  {\bibinfo  {journal} {Cells and Development}\ }\textbf {\bibinfo {volume}
  {168}},\ \bibinfo {pages} {203727} (\bibinfo {year} {2021})}\BibitemShut
  {NoStop}%
\bibitem [{\citenamefont {Park}\ \emph {et~al.}(2015)\citenamefont {Park},
  \citenamefont {Kim}, \citenamefont {Bi}, \citenamefont {Mitchel},
  \citenamefont {Qazvini}, \citenamefont {Tantisira}, \citenamefont {Park},
  \citenamefont {McGill}, \citenamefont {Kim}, \citenamefont {Gweon},
  \citenamefont {Notbohm}, \citenamefont {Steward~Jr}, \citenamefont {Burger},
  \citenamefont {Randell}, \citenamefont {Kho}, \citenamefont {Tambe},
  \citenamefont {Hardin}, \citenamefont {Shore}, \citenamefont {Israel},
  \citenamefont {Weitz}, \citenamefont {Tschumperlin}, \citenamefont {Henske},
  \citenamefont {Weiss}, \citenamefont {Manning}, \citenamefont {Butler},
  \citenamefont {Drazen},\ and\ \citenamefont {Fredberg}}]{park2015a}%
  \BibitemOpen
  \bibfield  {author} {\bibinfo {author} {\bibfnamefont {J.-A.}\ \bibnamefont
  {Park}}, \bibinfo {author} {\bibfnamefont {J.~H.}\ \bibnamefont {Kim}},
  \bibinfo {author} {\bibfnamefont {D.}~\bibnamefont {Bi}}, \bibinfo {author}
  {\bibfnamefont {J.~A.}\ \bibnamefont {Mitchel}}, \bibinfo {author}
  {\bibfnamefont {N.~T.}\ \bibnamefont {Qazvini}}, \bibinfo {author}
  {\bibfnamefont {K.}~\bibnamefont {Tantisira}}, \bibinfo {author}
  {\bibfnamefont {C.~Y.}\ \bibnamefont {Park}}, \bibinfo {author}
  {\bibfnamefont {M.}~\bibnamefont {McGill}}, \bibinfo {author} {\bibfnamefont
  {S.-H.}\ \bibnamefont {Kim}}, \bibinfo {author} {\bibfnamefont
  {B.}~\bibnamefont {Gweon}}, \bibinfo {author} {\bibfnamefont
  {J.}~\bibnamefont {Notbohm}}, \bibinfo {author} {\bibfnamefont
  {R.}~\bibnamefont {Steward~Jr}}, \bibinfo {author} {\bibfnamefont
  {S.}~\bibnamefont {Burger}}, \bibinfo {author} {\bibfnamefont {S.~H.}\
  \bibnamefont {Randell}}, \bibinfo {author} {\bibfnamefont {A.~T.}\
  \bibnamefont {Kho}}, \bibinfo {author} {\bibfnamefont {D.~T.}\ \bibnamefont
  {Tambe}}, \bibinfo {author} {\bibfnamefont {C.}~\bibnamefont {Hardin}},
  \bibinfo {author} {\bibfnamefont {S.~A.}\ \bibnamefont {Shore}}, \bibinfo
  {author} {\bibfnamefont {E.}~\bibnamefont {Israel}}, \bibinfo {author}
  {\bibfnamefont {D.~A.}\ \bibnamefont {Weitz}}, \bibinfo {author}
  {\bibfnamefont {D.~J.}\ \bibnamefont {Tschumperlin}}, \bibinfo {author}
  {\bibfnamefont {E.}~\bibnamefont {Henske}}, \bibinfo {author} {\bibfnamefont
  {S.~T.}\ \bibnamefont {Weiss}}, \bibinfo {author} {\bibfnamefont {M.~L.}\
  \bibnamefont {Manning}}, \bibinfo {author} {\bibfnamefont {J.~P.}\
  \bibnamefont {Butler}}, \bibinfo {author} {\bibfnamefont {J.~M.}\
  \bibnamefont {Drazen}}, \ and\ \bibinfo {author} {\bibfnamefont {J.~J.}\
  \bibnamefont {Fredberg}},\ }\href {\doibase 10.1038/nmat4357} {\bibfield
  {journal} {\bibinfo  {journal} {Nat. Mat.}\ }\textbf {\bibinfo {volume}
  {14}},\ \bibinfo {pages} {1040} (\bibinfo {year} {2015})}\BibitemShut
  {NoStop}%
\bibitem [{\citenamefont {Angelini}\ \emph {et~al.}(2011)\citenamefont
  {Angelini}, \citenamefont {Hannezo}, \citenamefont {Trepat}, \citenamefont
  {Marquez}, \citenamefont {Fredberg},\ and\ \citenamefont
  {Weitz}}]{angelini2011}%
  \BibitemOpen
  \bibfield  {author} {\bibinfo {author} {\bibfnamefont {T.~E.}\ \bibnamefont
  {Angelini}}, \bibinfo {author} {\bibfnamefont {E.}~\bibnamefont {Hannezo}},
  \bibinfo {author} {\bibfnamefont {X.}~\bibnamefont {Trepat}}, \bibinfo
  {author} {\bibfnamefont {M.}~\bibnamefont {Marquez}}, \bibinfo {author}
  {\bibfnamefont {J.~J.}\ \bibnamefont {Fredberg}}, \ and\ \bibinfo {author}
  {\bibfnamefont {D.~A.}\ \bibnamefont {Weitz}},\ }\href {\doibase
  10.1073/pnas.1010059108} {\bibfield  {journal} {\bibinfo  {journal} {Proc.
  Nat. Acad. Sci. (USA)}\ }\textbf {\bibinfo {volume} {108}},\ \bibinfo {pages}
  {4714} (\bibinfo {year} {2011})}\BibitemShut {NoStop}%
\bibitem [{\citenamefont {Sadhukhan}\ \emph
  {et~al.}(2024{\natexlab{a}})\citenamefont {Sadhukhan}, \citenamefont {Dey},
  \citenamefont {Karmakar},\ and\ \citenamefont {Nandi}}]{activereview}%
  \BibitemOpen
  \bibfield  {author} {\bibinfo {author} {\bibfnamefont {S.}~\bibnamefont
  {Sadhukhan}}, \bibinfo {author} {\bibfnamefont {S.}~\bibnamefont {Dey}},
  \bibinfo {author} {\bibfnamefont {S.}~\bibnamefont {Karmakar}}, \ and\
  \bibinfo {author} {\bibfnamefont {S.~K.}\ \bibnamefont {Nandi}},\ }\href
  {\doibase 10.1140/ep js/s11734-024-01188-1} {\bibfield  {journal} {\bibinfo
  {journal} {Eur. Phys. J. Spec. Top.}\ } (\bibinfo {year}
  {2024}{\natexlab{a}}),\ 10.1140/ep js/s11734-024-01188-1}\BibitemShut
  {NoStop}%
\bibitem [{\citenamefont {Vishwakarma}\ \emph {et~al.}(2020)\citenamefont
  {Vishwakarma}, \citenamefont {Thurakkal}, \citenamefont {Spatz},\ and\
  \citenamefont {Das}}]{vishwakarma2020}%
  \BibitemOpen
  \bibfield  {author} {\bibinfo {author} {\bibfnamefont {M.}~\bibnamefont
  {Vishwakarma}}, \bibinfo {author} {\bibfnamefont {B.}~\bibnamefont
  {Thurakkal}}, \bibinfo {author} {\bibfnamefont {J.~P.}\ \bibnamefont
  {Spatz}}, \ and\ \bibinfo {author} {\bibfnamefont {T.}~\bibnamefont {Das}},\
  }\href {\doibase 10.1098/rstb.2019.0391} {\bibfield  {journal} {\bibinfo
  {journal} {Phil. Tran. R. Soc. B: Biol. Sci.}\ }\textbf {\bibinfo {volume}
  {375}},\ \bibinfo {pages} {20190391} (\bibinfo {year} {2020})}\BibitemShut
  {NoStop}%
\bibitem [{\citenamefont {Poujade}\ \emph {et~al.}(2007)\citenamefont
  {Poujade}, \citenamefont {Grasland-Mongrain}, \citenamefont {Hertzog},
  \citenamefont {Jouanneau}, \citenamefont {Chavrier}, \citenamefont {Ladoux},
  \citenamefont {Buguin},\ and\ \citenamefont {Silberzan}}]{poujade2007}%
  \BibitemOpen
  \bibfield  {author} {\bibinfo {author} {\bibfnamefont {M.}~\bibnamefont
  {Poujade}}, \bibinfo {author} {\bibfnamefont {E.}~\bibnamefont
  {Grasland-Mongrain}}, \bibinfo {author} {\bibfnamefont {A.}~\bibnamefont
  {Hertzog}}, \bibinfo {author} {\bibfnamefont {J.}~\bibnamefont {Jouanneau}},
  \bibinfo {author} {\bibfnamefont {P.}~\bibnamefont {Chavrier}}, \bibinfo
  {author} {\bibfnamefont {B.}~\bibnamefont {Ladoux}}, \bibinfo {author}
  {\bibfnamefont {A.}~\bibnamefont {Buguin}}, \ and\ \bibinfo {author}
  {\bibfnamefont {P.}~\bibnamefont {Silberzan}},\ }\href {\doibase
  10.1073/pnas.0705062104} {\bibfield  {journal} {\bibinfo  {journal} {Proc.
  Natl. Acad. Sci. (USA)}\ }\textbf {\bibinfo {volume} {104}},\ \bibinfo
  {pages} {15988} (\bibinfo {year} {2007})}\BibitemShut {NoStop}%
\bibitem [{\citenamefont {Tetley}\ and\ \citenamefont
  {Mao}(2018)}]{tetley2018}%
  \BibitemOpen
  \bibfield  {author} {\bibinfo {author} {\bibfnamefont {R.~J.}\ \bibnamefont
  {Tetley}}\ and\ \bibinfo {author} {\bibfnamefont {Y.}~\bibnamefont {Mao}},\
  }\href {\doibase 10.1098/rstb.2017.0328} {\bibfield  {journal} {\bibinfo
  {journal} {Philosophical Transactions of the Royal Society B: Biological
  Sciences}\ }\textbf {\bibinfo {volume} {373}},\ \bibinfo {pages} {20170328}
  (\bibinfo {year} {2018})}\BibitemShut {NoStop}%
\bibitem [{\citenamefont {Spurlin}\ \emph {et~al.}(2019)\citenamefont
  {Spurlin}, \citenamefont {Siedlik}, \citenamefont {Nerger}, \citenamefont
  {Pang}, \citenamefont {Jayaraman}, \citenamefont {Zhang},\ and\ \citenamefont
  {Nelson}}]{spurlin2019}%
  \BibitemOpen
  \bibfield  {author} {\bibinfo {author} {\bibfnamefont {J.~W.}\ \bibnamefont
  {Spurlin}}, \bibinfo {author} {\bibfnamefont {M.~J.}\ \bibnamefont
  {Siedlik}}, \bibinfo {author} {\bibfnamefont {B.~A.}\ \bibnamefont {Nerger}},
  \bibinfo {author} {\bibfnamefont {M.-F.}\ \bibnamefont {Pang}}, \bibinfo
  {author} {\bibfnamefont {S.}~\bibnamefont {Jayaraman}}, \bibinfo {author}
  {\bibfnamefont {R.}~\bibnamefont {Zhang}}, \ and\ \bibinfo {author}
  {\bibfnamefont {C.~M.}\ \bibnamefont {Nelson}},\ }\href {\doibase
  10.1242/dev.175257} {\bibfield  {journal} {\bibinfo  {journal} {Development}\
  }\textbf {\bibinfo {volume} {146}} (\bibinfo {year} {2019}),\
  10.1242/dev.175257}\BibitemShut {NoStop}%
\bibitem [{\citenamefont {Mongera}\ \emph {et~al.}(2018)\citenamefont
  {Mongera}, \citenamefont {Rowghanian}, \citenamefont {Gustafson},
  \citenamefont {Shelton}, \citenamefont {Kealhofer}, \citenamefont {Carn},
  \citenamefont {Serwane}, \citenamefont {Lucio}, \citenamefont {Giammona},\
  and\ \citenamefont {Camp{\`a}s}}]{mongera2018}%
  \BibitemOpen
  \bibfield  {author} {\bibinfo {author} {\bibfnamefont {A.}~\bibnamefont
  {Mongera}}, \bibinfo {author} {\bibfnamefont {P.}~\bibnamefont {Rowghanian}},
  \bibinfo {author} {\bibfnamefont {H.~J.}\ \bibnamefont {Gustafson}}, \bibinfo
  {author} {\bibfnamefont {E.}~\bibnamefont {Shelton}}, \bibinfo {author}
  {\bibfnamefont {D.~A.}\ \bibnamefont {Kealhofer}}, \bibinfo {author}
  {\bibfnamefont {E.~K.}\ \bibnamefont {Carn}}, \bibinfo {author}
  {\bibfnamefont {F.}~\bibnamefont {Serwane}}, \bibinfo {author} {\bibfnamefont
  {A.~A.}\ \bibnamefont {Lucio}}, \bibinfo {author} {\bibfnamefont
  {J.}~\bibnamefont {Giammona}}, \ and\ \bibinfo {author} {\bibfnamefont
  {O.}~\bibnamefont {Camp{\`a}s}},\ }\href {\doibase 10.1038/s41586-018-0479-2}
  {\bibfield  {journal} {\bibinfo  {journal} {Nature}\ }\textbf {\bibinfo
  {volume} {561}},\ \bibinfo {pages} {401} (\bibinfo {year}
  {2018})}\BibitemShut {NoStop}%
\bibitem [{\citenamefont {Friedl}\ and\ \citenamefont
  {Wolf}(2003)}]{friedl2003a}%
  \BibitemOpen
  \bibfield  {author} {\bibinfo {author} {\bibfnamefont {P.}~\bibnamefont
  {Friedl}}\ and\ \bibinfo {author} {\bibfnamefont {K.}~\bibnamefont {Wolf}},\
  }\href {\doibase 10.1038/nrc1075} {\bibfield  {journal} {\bibinfo  {journal}
  {Nat. Rev. Cancer}\ }\textbf {\bibinfo {volume} {3}},\ \bibinfo {pages} {362}
  (\bibinfo {year} {2003})}\BibitemShut {NoStop}%
\bibitem [{\citenamefont {Friedl}\ and\ \citenamefont
  {Gilmour}(2009)}]{friedl2009b}%
  \BibitemOpen
  \bibfield  {author} {\bibinfo {author} {\bibfnamefont {P.}~\bibnamefont
  {Friedl}}\ and\ \bibinfo {author} {\bibfnamefont {D.}~\bibnamefont
  {Gilmour}},\ }\href {\doibase 10.1038/nrm2720} {\bibfield  {journal}
  {\bibinfo  {journal} {Nat. Rev. Mol. Cell Biol.}\ }\textbf {\bibinfo {volume}
  {10}},\ \bibinfo {pages} {445} (\bibinfo {year} {2009})}\BibitemShut
  {NoStop}%
\bibitem [{\citenamefont {Garcia}\ \emph {et~al.}(2015)\citenamefont {Garcia},
  \citenamefont {Hannezo}, \citenamefont {Elgeti}, \citenamefont {Joanny},
  \citenamefont {Silberzan},\ and\ \citenamefont {Gov}}]{garcia2015}%
  \BibitemOpen
  \bibfield  {author} {\bibinfo {author} {\bibfnamefont {S.}~\bibnamefont
  {Garcia}}, \bibinfo {author} {\bibfnamefont {E.}~\bibnamefont {Hannezo}},
  \bibinfo {author} {\bibfnamefont {J.}~\bibnamefont {Elgeti}}, \bibinfo
  {author} {\bibfnamefont {J.-F.}\ \bibnamefont {Joanny}}, \bibinfo {author}
  {\bibfnamefont {P.}~\bibnamefont {Silberzan}}, \ and\ \bibinfo {author}
  {\bibfnamefont {N.~S.}\ \bibnamefont {Gov}},\ }\href {\doibase
  10.1073/pnas.1510973112} {\bibfield  {journal} {\bibinfo  {journal} {Proc.
  Nat. Acad. Sci. (USA)}\ }\textbf {\bibinfo {volume} {112}},\ \bibinfo {pages}
  {15314} (\bibinfo {year} {2015})}\BibitemShut {NoStop}%
\bibitem [{\citenamefont {Malinverno}\ \emph {et~al.}(2017)\citenamefont
  {Malinverno}, \citenamefont {Corallino}, \citenamefont {Giavazzi},
  \citenamefont {Bergert}, \citenamefont {Li}, \citenamefont {Leoni},
  \citenamefont {Disanza}, \citenamefont {Frittoli}, \citenamefont {Oldani},
  \citenamefont {Martini}, \citenamefont {Lendenmann}, \citenamefont
  {Deflorian}, \citenamefont {Beznoussenko}, \citenamefont {Poulikakos},
  \citenamefont {Ong}, \citenamefont {Uroz}, \citenamefont {Trepat},
  \citenamefont {Parazzoli}, \citenamefont {Maiuri}, \citenamefont {Yu},
  \citenamefont {Ferrari}, \citenamefont {Cerbino},\ and\ \citenamefont
  {Scita}}]{malinverno2017}%
  \BibitemOpen
  \bibfield  {author} {\bibinfo {author} {\bibfnamefont {C.}~\bibnamefont
  {Malinverno}}, \bibinfo {author} {\bibfnamefont {S.}~\bibnamefont
  {Corallino}}, \bibinfo {author} {\bibfnamefont {F.}~\bibnamefont {Giavazzi}},
  \bibinfo {author} {\bibfnamefont {M.}~\bibnamefont {Bergert}}, \bibinfo
  {author} {\bibfnamefont {Q.}~\bibnamefont {Li}}, \bibinfo {author}
  {\bibfnamefont {M.}~\bibnamefont {Leoni}}, \bibinfo {author} {\bibfnamefont
  {A.}~\bibnamefont {Disanza}}, \bibinfo {author} {\bibfnamefont
  {E.}~\bibnamefont {Frittoli}}, \bibinfo {author} {\bibfnamefont
  {A.}~\bibnamefont {Oldani}}, \bibinfo {author} {\bibfnamefont
  {E.}~\bibnamefont {Martini}}, \bibinfo {author} {\bibfnamefont
  {T.}~\bibnamefont {Lendenmann}}, \bibinfo {author} {\bibfnamefont
  {G.}~\bibnamefont {Deflorian}}, \bibinfo {author} {\bibfnamefont {G.~V.}\
  \bibnamefont {Beznoussenko}}, \bibinfo {author} {\bibfnamefont
  {D.}~\bibnamefont {Poulikakos}}, \bibinfo {author} {\bibfnamefont {K.~H.}\
  \bibnamefont {Ong}}, \bibinfo {author} {\bibfnamefont {M.}~\bibnamefont
  {Uroz}}, \bibinfo {author} {\bibfnamefont {X.}~\bibnamefont {Trepat}},
  \bibinfo {author} {\bibfnamefont {D.}~\bibnamefont {Parazzoli}}, \bibinfo
  {author} {\bibfnamefont {P.}~\bibnamefont {Maiuri}}, \bibinfo {author}
  {\bibfnamefont {W.}~\bibnamefont {Yu}}, \bibinfo {author} {\bibfnamefont
  {A.}~\bibnamefont {Ferrari}}, \bibinfo {author} {\bibfnamefont
  {R.}~\bibnamefont {Cerbino}}, \ and\ \bibinfo {author} {\bibfnamefont
  {G.}~\bibnamefont {Scita}},\ }\href {\doibase 10.1038/nmat4848} {\bibfield
  {journal} {\bibinfo  {journal} {Nat. Mat.}\ }\textbf {\bibinfo {volume}
  {16}},\ \bibinfo {pages} {587} (\bibinfo {year} {2017})}\BibitemShut
  {NoStop}%
\bibitem [{\citenamefont {Palamidessi}\ \emph {et~al.}(2019)\citenamefont
  {Palamidessi}, \citenamefont {Malinverno}, \citenamefont {Frittoli},
  \citenamefont {Corallino}, \citenamefont {Barbieri}, \citenamefont
  {Sigismund}, \citenamefont {Beznoussenko}, \citenamefont {Martini},
  \citenamefont {Garre}, \citenamefont {Ferrara}, \citenamefont {Tripodo},
  \citenamefont {Ascione}, \citenamefont {Cavalcanti-Adam}, \citenamefont {Li},
  \citenamefont {Di~Fiore}, \citenamefont {Parazzoli}, \citenamefont
  {Giavazzi}, \citenamefont {Cerbino},\ and\ \citenamefont
  {Scita}}]{palamidessi2019}%
  \BibitemOpen
  \bibfield  {author} {\bibinfo {author} {\bibfnamefont {A.}~\bibnamefont
  {Palamidessi}}, \bibinfo {author} {\bibfnamefont {C.}~\bibnamefont
  {Malinverno}}, \bibinfo {author} {\bibfnamefont {E.}~\bibnamefont
  {Frittoli}}, \bibinfo {author} {\bibfnamefont {S.}~\bibnamefont {Corallino}},
  \bibinfo {author} {\bibfnamefont {E.}~\bibnamefont {Barbieri}}, \bibinfo
  {author} {\bibfnamefont {S.}~\bibnamefont {Sigismund}}, \bibinfo {author}
  {\bibfnamefont {G.~V.}\ \bibnamefont {Beznoussenko}}, \bibinfo {author}
  {\bibfnamefont {E.}~\bibnamefont {Martini}}, \bibinfo {author} {\bibfnamefont
  {M.}~\bibnamefont {Garre}}, \bibinfo {author} {\bibfnamefont
  {I.}~\bibnamefont {Ferrara}}, \bibinfo {author} {\bibfnamefont
  {C.}~\bibnamefont {Tripodo}}, \bibinfo {author} {\bibfnamefont
  {F.}~\bibnamefont {Ascione}}, \bibinfo {author} {\bibfnamefont {E.~A.}\
  \bibnamefont {Cavalcanti-Adam}}, \bibinfo {author} {\bibfnamefont
  {Q.}~\bibnamefont {Li}}, \bibinfo {author} {\bibfnamefont {P.~P.}\
  \bibnamefont {Di~Fiore}}, \bibinfo {author} {\bibfnamefont {D.}~\bibnamefont
  {Parazzoli}}, \bibinfo {author} {\bibfnamefont {F.}~\bibnamefont {Giavazzi}},
  \bibinfo {author} {\bibfnamefont {R.}~\bibnamefont {Cerbino}}, \ and\
  \bibinfo {author} {\bibfnamefont {G.}~\bibnamefont {Scita}},\ }\href
  {\doibase 10.1038/s41563-019-0425-1} {\bibfield  {journal} {\bibinfo
  {journal} {Nat. Mat.}\ }\textbf {\bibinfo {volume} {18}},\ \bibinfo {pages}
  {1252—1263} (\bibinfo {year} {2019})}\BibitemShut {NoStop}%
\bibitem [{\citenamefont {Giavazzi}\ \emph {et~al.}(2017)\citenamefont
  {Giavazzi}, \citenamefont {Malinverno}, \citenamefont {Corallino},
  \citenamefont {Ginelli}, \citenamefont {Scita},\ and\ \citenamefont
  {Cerbino}}]{giavazzi2017}%
  \BibitemOpen
  \bibfield  {author} {\bibinfo {author} {\bibfnamefont {F.}~\bibnamefont
  {Giavazzi}}, \bibinfo {author} {\bibfnamefont {C.}~\bibnamefont
  {Malinverno}}, \bibinfo {author} {\bibfnamefont {S.}~\bibnamefont
  {Corallino}}, \bibinfo {author} {\bibfnamefont {F.}~\bibnamefont {Ginelli}},
  \bibinfo {author} {\bibfnamefont {G.}~\bibnamefont {Scita}}, \ and\ \bibinfo
  {author} {\bibfnamefont {R.}~\bibnamefont {Cerbino}},\ }\href {\doibase
  10.1088/1361-6463/aa7f8e} {\bibfield  {journal} {\bibinfo  {journal} {J.
  Phys. D: Appl. Phys.}\ }\textbf {\bibinfo {volume} {50}},\ \bibinfo {pages}
  {384003} (\bibinfo {year} {2017})}\BibitemShut {NoStop}%
\bibitem [{\citenamefont {Berthier}\ and\ \citenamefont
  {Biroli}(2011)}]{berthier2011a}%
  \BibitemOpen
  \bibfield  {author} {\bibinfo {author} {\bibfnamefont {L.}~\bibnamefont
  {Berthier}}\ and\ \bibinfo {author} {\bibfnamefont {G.}~\bibnamefont
  {Biroli}},\ }\href {\doibase 10.1103/RevModPhys.83.587} {\bibfield  {journal}
  {\bibinfo  {journal} {Rev. Mod. Phys.}\ }\textbf {\bibinfo {volume} {83}},\
  \bibinfo {pages} {587} (\bibinfo {year} {2011})}\BibitemShut {NoStop}%
\bibitem [{\citenamefont {Berthier}\ \emph {et~al.}(2019)\citenamefont
  {Berthier}, \citenamefont {Flenner},\ and\ \citenamefont
  {Szamel}}]{berthier2019}%
  \BibitemOpen
  \bibfield  {author} {\bibinfo {author} {\bibfnamefont {L.}~\bibnamefont
  {Berthier}}, \bibinfo {author} {\bibfnamefont {E.}~\bibnamefont {Flenner}}, \
  and\ \bibinfo {author} {\bibfnamefont {G.}~\bibnamefont {Szamel}},\ }\href
  {\doibase 10.1063/1.5093240} {\bibfield  {journal} {\bibinfo  {journal} {J.
  Chem. Phys.}\ }\textbf {\bibinfo {volume} {150}},\ \bibinfo {pages} {200901}
  (\bibinfo {year} {2019})}\BibitemShut {NoStop}%
\bibitem [{\citenamefont {Bi}\ \emph {et~al.}(2016)\citenamefont {Bi},
  \citenamefont {Yang}, \citenamefont {Marchetti},\ and\ \citenamefont
  {Manning}}]{bi2016}%
  \BibitemOpen
  \bibfield  {author} {\bibinfo {author} {\bibfnamefont {D.}~\bibnamefont
  {Bi}}, \bibinfo {author} {\bibfnamefont {X.}~\bibnamefont {Yang}}, \bibinfo
  {author} {\bibfnamefont {M.~C.}\ \bibnamefont {Marchetti}}, \ and\ \bibinfo
  {author} {\bibfnamefont {M.~L.}\ \bibnamefont {Manning}},\ }\href {\doibase
  10.1103/PhysRevX.6.021011} {\bibfield  {journal} {\bibinfo  {journal} {Phys.
  Rev. X}\ }\textbf {\bibinfo {volume} {6}},\ \bibinfo {pages} {021011}
  (\bibinfo {year} {2016})}\BibitemShut {NoStop}%
\bibitem [{\citenamefont {Li}\ \emph {et~al.}(2021)\citenamefont {Li},
  \citenamefont {Wei}, \citenamefont {Paoluzzi},\ and\ \citenamefont
  {Ciamarra}}]{li2021}%
  \BibitemOpen
  \bibfield  {author} {\bibinfo {author} {\bibfnamefont {Y.-W.}\ \bibnamefont
  {Li}}, \bibinfo {author} {\bibfnamefont {L.~L.~Y.}\ \bibnamefont {Wei}},
  \bibinfo {author} {\bibfnamefont {M.}~\bibnamefont {Paoluzzi}}, \ and\
  \bibinfo {author} {\bibfnamefont {M.~P.}\ \bibnamefont {Ciamarra}},\ }\href
  {\doibase 10.1103/PhysRevE.103.022607} {\bibfield  {journal} {\bibinfo
  {journal} {Phys. Rev. E}\ }\textbf {\bibinfo {volume} {103}},\ \bibinfo
  {pages} {022607} (\bibinfo {year} {2021})}\BibitemShut {NoStop}%
\bibitem [{\citenamefont {Atia}\ \emph {et~al.}(2018)\citenamefont {Atia},
  \citenamefont {Bi}, \citenamefont {Sharma}, \citenamefont {Mitchel},
  \citenamefont {Gweon}, \citenamefont {A.~Koehler}, \citenamefont {DeCamp},
  \citenamefont {Lan}, \citenamefont {Kim}, \citenamefont {Hirsch},
  \citenamefont {Pegoraro}, \citenamefont {Lee}, \citenamefont {Starr},
  \citenamefont {Weitz}, \citenamefont {Martin}, \citenamefont {Park},
  \citenamefont {Butler},\ and\ \citenamefont {Fredberg}}]{atia2018}%
  \BibitemOpen
  \bibfield  {author} {\bibinfo {author} {\bibfnamefont {L.}~\bibnamefont
  {Atia}}, \bibinfo {author} {\bibfnamefont {D.}~\bibnamefont {Bi}}, \bibinfo
  {author} {\bibfnamefont {Y.}~\bibnamefont {Sharma}}, \bibinfo {author}
  {\bibfnamefont {J.~A.}\ \bibnamefont {Mitchel}}, \bibinfo {author}
  {\bibfnamefont {B.}~\bibnamefont {Gweon}}, \bibinfo {author} {\bibfnamefont
  {S.}~\bibnamefont {A.~Koehler}}, \bibinfo {author} {\bibfnamefont {S.~J.}\
  \bibnamefont {DeCamp}}, \bibinfo {author} {\bibfnamefont {B.}~\bibnamefont
  {Lan}}, \bibinfo {author} {\bibfnamefont {J.~H.}\ \bibnamefont {Kim}},
  \bibinfo {author} {\bibfnamefont {R.}~\bibnamefont {Hirsch}}, \bibinfo
  {author} {\bibfnamefont {A.~F.}\ \bibnamefont {Pegoraro}}, \bibinfo {author}
  {\bibfnamefont {K.~H.}\ \bibnamefont {Lee}}, \bibinfo {author} {\bibfnamefont
  {J.~R.}\ \bibnamefont {Starr}}, \bibinfo {author} {\bibfnamefont {D.~A.}\
  \bibnamefont {Weitz}}, \bibinfo {author} {\bibfnamefont {A.~C.}\ \bibnamefont
  {Martin}}, \bibinfo {author} {\bibfnamefont {J.-A.}\ \bibnamefont {Park}},
  \bibinfo {author} {\bibfnamefont {J.~P.}\ \bibnamefont {Butler}}, \ and\
  \bibinfo {author} {\bibfnamefont {J.~J.}\ \bibnamefont {Fredberg}},\ }\href
  {\doibase 10.1038/s41567-018-0089-9} {\bibfield  {journal} {\bibinfo
  {journal} {Nat. Phys.}\ }\textbf {\bibinfo {volume} {14}},\ \bibinfo {pages}
  {613} (\bibinfo {year} {2018})}\BibitemShut {NoStop}%
\bibitem [{\citenamefont {Sadhukhan}\ and\ \citenamefont
  {Nandi}(2021)}]{sadhukhan2021a}%
  \BibitemOpen
  \bibfield  {author} {\bibinfo {author} {\bibfnamefont {S.}~\bibnamefont
  {Sadhukhan}}\ and\ \bibinfo {author} {\bibfnamefont {S.~K.}\ \bibnamefont
  {Nandi}},\ }\href {\doibase 10.1103/PhysRevE.103.062403} {\bibfield
  {journal} {\bibinfo  {journal} {Phys. Rev. E}\ }\textbf {\bibinfo {volume}
  {103}},\ \bibinfo {pages} {062403} (\bibinfo {year} {2021})}\BibitemShut
  {NoStop}%
\bibitem [{\citenamefont {Sadhukhan}\ and\ \citenamefont
  {Nandi}(2022)}]{sadhukhan2022}%
  \BibitemOpen
  \bibfield  {author} {\bibinfo {author} {\bibfnamefont {S.}~\bibnamefont
  {Sadhukhan}}\ and\ \bibinfo {author} {\bibfnamefont {S.~K.}\ \bibnamefont
  {Nandi}},\ }\href {\doibase 10.7554/eLife.76406} {\bibfield  {journal}
  {\bibinfo  {journal} {eLife}\ }\textbf {\bibinfo {volume} {11}},\ \bibinfo
  {pages} {e76406} (\bibinfo {year} {2022})}\BibitemShut {NoStop}%
\bibitem [{\citenamefont {Sadhukhan}\ \emph
  {et~al.}(2024{\natexlab{b}})\citenamefont {Sadhukhan}, \citenamefont {Nandi},
  \citenamefont {Pandey}, \citenamefont {Paoluzzi}, \citenamefont {Dasgupta},
  \citenamefont {Gov},\ and\ \citenamefont {Nandi}}]{sadhukhan2024}%
  \BibitemOpen
  \bibfield  {author} {\bibinfo {author} {\bibfnamefont {S.}~\bibnamefont
  {Sadhukhan}}, \bibinfo {author} {\bibfnamefont {M.~K.}\ \bibnamefont
  {Nandi}}, \bibinfo {author} {\bibfnamefont {S.}~\bibnamefont {Pandey}},
  \bibinfo {author} {\bibfnamefont {M.}~\bibnamefont {Paoluzzi}}, \bibinfo
  {author} {\bibfnamefont {C.}~\bibnamefont {Dasgupta}}, \bibinfo {author}
  {\bibfnamefont {N.}~\bibnamefont {Gov}}, \ and\ \bibinfo {author}
  {\bibfnamefont {S.~K.}\ \bibnamefont {Nandi}},\ }\href@noop {} {\bibfield
  {journal} {\bibinfo  {journal} {arXiv}\ ,\ \bibinfo {pages} {2403.08437}}
  (\bibinfo {year} {2024}{\natexlab{b}})}\BibitemShut {NoStop}%
\bibitem [{\citenamefont {Arora}\ \emph {et~al.}(2024)\citenamefont {Arora},
  \citenamefont {Sadhukhan}, \citenamefont {Nandi}, \citenamefont {Bi},
  \citenamefont {Sood},\ and\ \citenamefont {Ganapathy}}]{arora2024}%
  \BibitemOpen
  \bibfield  {author} {\bibinfo {author} {\bibfnamefont {P.}~\bibnamefont
  {Arora}}, \bibinfo {author} {\bibfnamefont {S.}~\bibnamefont {Sadhukhan}},
  \bibinfo {author} {\bibfnamefont {S.~K.}\ \bibnamefont {Nandi}}, \bibinfo
  {author} {\bibfnamefont {D.}~\bibnamefont {Bi}}, \bibinfo {author}
  {\bibfnamefont {A.~K.}\ \bibnamefont {Sood}}, \ and\ \bibinfo {author}
  {\bibfnamefont {R.}~\bibnamefont {Ganapathy}},\ }\href {\doibase
  10.48550/arXiv.2401.13437} {\bibfield  {journal} {\bibinfo  {journal}
  {arXiv}\ ,\ \bibinfo {pages} {2401.13437}} (\bibinfo {year}
  {2024})}\BibitemShut {NoStop}%
\bibitem [{\citenamefont {Sussman}\ \emph {et~al.}(2018)\citenamefont
  {Sussman}, \citenamefont {Paoluzzi}, \citenamefont {Marchetti},\ and\
  \citenamefont {Manning}}]{sussman2018b}%
  \BibitemOpen
  \bibfield  {author} {\bibinfo {author} {\bibfnamefont {D.~M.}\ \bibnamefont
  {Sussman}}, \bibinfo {author} {\bibfnamefont {M.}~\bibnamefont {Paoluzzi}},
  \bibinfo {author} {\bibfnamefont {M.~C.}\ \bibnamefont {Marchetti}}, \ and\
  \bibinfo {author} {\bibfnamefont {M.~L.}\ \bibnamefont {Manning}},\ }\href
  {\doibase 10.1209/0295-5075/121/36001} {\bibfield  {journal} {\bibinfo
  {journal} {Europhys. Lett.}\ }\textbf {\bibinfo {volume} {121}},\ \bibinfo
  {pages} {36001} (\bibinfo {year} {2018})}\BibitemShut {NoStop}%
\bibitem [{\citenamefont {Graner}\ and\ \citenamefont
  {Glazier}(1992)}]{graner1992}%
  \BibitemOpen
  \bibfield  {author} {\bibinfo {author} {\bibfnamefont {F.}~\bibnamefont
  {Graner}}\ and\ \bibinfo {author} {\bibfnamefont {J.~A.}\ \bibnamefont
  {Glazier}},\ }\href {\doibase 10.1103/PhysRevLett.69.2013} {\bibfield
  {journal} {\bibinfo  {journal} {Phys. Rev. Lett.}\ }\textbf {\bibinfo
  {volume} {69}},\ \bibinfo {pages} {2013} (\bibinfo {year}
  {1992})}\BibitemShut {NoStop}%
\bibitem [{\citenamefont {Hogeweg}(2000)}]{hogeweg2000}%
  \BibitemOpen
  \bibfield  {author} {\bibinfo {author} {\bibfnamefont {P.}~\bibnamefont
  {Hogeweg}},\ }\href {\doibase 10.1006/jtbi.2000.1087} {\bibfield  {journal}
  {\bibinfo  {journal} {J. Theor. Biol.}\ }\textbf {\bibinfo {volume} {203}},\
  \bibinfo {pages} {317} (\bibinfo {year} {2000})}\BibitemShut {NoStop}%
\bibitem [{\citenamefont {Honda}\ and\ \citenamefont
  {Eguchi}(1980)}]{honda1980}%
  \BibitemOpen
  \bibfield  {author} {\bibinfo {author} {\bibfnamefont {H.}~\bibnamefont
  {Honda}}\ and\ \bibinfo {author} {\bibfnamefont {G.}~\bibnamefont {Eguchi}},\
  }\href {\doibase 10.1016/S0022-5193(80)80021-X} {\bibfield  {journal}
  {\bibinfo  {journal} {J. Theor. Biol.}\ }\textbf {\bibinfo {volume} {84}},\
  \bibinfo {pages} {575} (\bibinfo {year} {1980})}\BibitemShut {NoStop}%
\bibitem [{\citenamefont {Farhadifar}\ \emph {et~al.}(2007)\citenamefont
  {Farhadifar}, \citenamefont {R{\"{o}}per}, \citenamefont {Aigouy},
  \citenamefont {Eaton},\ and\ \citenamefont
  {J{\"{u}}licher}}]{farhadifar2007}%
  \BibitemOpen
  \bibfield  {author} {\bibinfo {author} {\bibfnamefont {R.}~\bibnamefont
  {Farhadifar}}, \bibinfo {author} {\bibfnamefont {J.-C.}\ \bibnamefont
  {R{\"{o}}per}}, \bibinfo {author} {\bibfnamefont {B.}~\bibnamefont {Aigouy}},
  \bibinfo {author} {\bibfnamefont {S.}~\bibnamefont {Eaton}}, \ and\ \bibinfo
  {author} {\bibfnamefont {F.}~\bibnamefont {J{\"{u}}licher}},\ }\href
  {\doibase 10.1016/j.cub.2007.11.049} {\bibfield  {journal} {\bibinfo
  {journal} {Curr. Biol.}\ }\textbf {\bibinfo {volume} {17}},\ \bibinfo {pages}
  {2095} (\bibinfo {year} {2007})}\BibitemShut {NoStop}%
\bibitem [{\citenamefont {Fletcher}\ \emph {et~al.}(2014)\citenamefont
  {Fletcher}, \citenamefont {Osterfield}, \citenamefont {Baker},\ and\
  \citenamefont {Shvartsman}}]{fletcher2014}%
  \BibitemOpen
  \bibfield  {author} {\bibinfo {author} {\bibfnamefont {A.}~\bibnamefont
  {Fletcher}}, \bibinfo {author} {\bibfnamefont {M.}~\bibnamefont
  {Osterfield}}, \bibinfo {author} {\bibfnamefont {R.}~\bibnamefont {Baker}}, \
  and\ \bibinfo {author} {\bibfnamefont {S.}~\bibnamefont {Shvartsman}},\
  }\href {\doibase https://doi.org/10.1016/j.bpj.2013.11.4498} {\bibfield
  {journal} {\bibinfo  {journal} {Biophys. J.}\ }\textbf {\bibinfo {volume}
  {106}},\ \bibinfo {pages} {2291} (\bibinfo {year} {2014})}\BibitemShut
  {NoStop}%
\bibitem [{\citenamefont {Albert}\ and\ \citenamefont
  {Schwarz}(2016)}]{albert2016}%
  \BibitemOpen
  \bibfield  {author} {\bibinfo {author} {\bibfnamefont {P.~J.}\ \bibnamefont
  {Albert}}\ and\ \bibinfo {author} {\bibfnamefont {U.~S.}\ \bibnamefont
  {Schwarz}},\ }\href {\doibase 10.1080/19336918.2016.1148864} {\bibfield
  {journal} {\bibinfo  {journal} {Cell Adhesion and Migration}\ }\textbf
  {\bibinfo {volume} {10}},\ \bibinfo {pages} {1} (\bibinfo {year}
  {2016})}\BibitemShut {NoStop}%
\bibitem [{\citenamefont {Prost}\ \emph {et~al.}(2015)\citenamefont {Prost},
  \citenamefont {J{\"{u}}licher},\ and\ \citenamefont {Joanny}}]{jacques2015}%
  \BibitemOpen
  \bibfield  {author} {\bibinfo {author} {\bibfnamefont {J.}~\bibnamefont
  {Prost}}, \bibinfo {author} {\bibfnamefont {F.}~\bibnamefont
  {J{\"{u}}licher}}, \ and\ \bibinfo {author} {\bibfnamefont {J.~F.}\
  \bibnamefont {Joanny}},\ }\href {\doibase 10.1038/nphys3224} {\bibfield
  {journal} {\bibinfo  {journal} {Nat. Phys.}\ }\textbf {\bibinfo {volume}
  {11}},\ \bibinfo {pages} {111} (\bibinfo {year} {2015})}\BibitemShut
  {NoStop}%
\bibitem [{\citenamefont {Bi}\ \emph {et~al.}(2014)\citenamefont {Bi},
  \citenamefont {Lopez}, \citenamefont {Schwarz},\ and\ \citenamefont
  {Manning}}]{bi2014}%
  \BibitemOpen
  \bibfield  {author} {\bibinfo {author} {\bibfnamefont {D.}~\bibnamefont
  {Bi}}, \bibinfo {author} {\bibfnamefont {J.~H.}\ \bibnamefont {Lopez}},
  \bibinfo {author} {\bibfnamefont {J.~M.}\ \bibnamefont {Schwarz}}, \ and\
  \bibinfo {author} {\bibfnamefont {M.~L.}\ \bibnamefont {Manning}},\ }\href
  {\doibase 10.1039/C3SM52893F} {\bibfield  {journal} {\bibinfo  {journal}
  {Soft Matter}\ }\textbf {\bibinfo {volume} {10}},\ \bibinfo {pages} {1885}
  (\bibinfo {year} {2014})}\BibitemShut {NoStop}%
\bibitem [{\citenamefont {Sussman}\ and\ \citenamefont
  {Merkel}(2018)}]{sussman2018a}%
  \BibitemOpen
  \bibfield  {author} {\bibinfo {author} {\bibfnamefont {D.~M.}\ \bibnamefont
  {Sussman}}\ and\ \bibinfo {author} {\bibfnamefont {M.}~\bibnamefont
  {Merkel}},\ }\href {\doibase 10.1039/C7SM02127E} {\bibfield  {journal}
  {\bibinfo  {journal} {Soft Matter}\ }\textbf {\bibinfo {volume} {14}},\
  \bibinfo {pages} {3397} (\bibinfo {year} {2018})}\BibitemShut {NoStop}%
\bibitem [{\citenamefont {Angell}(1991)}]{angell1991}%
  \BibitemOpen
  \bibfield  {author} {\bibinfo {author} {\bibfnamefont {C.}~\bibnamefont
  {Angell}},\ }\href {\doibase https://doi.org/10.1016/0022-3093(91)90266-9}
  {\bibfield  {journal} {\bibinfo  {journal} {J. Non-Crys. Solids}\ }\textbf
  {\bibinfo {volume} {131-133}},\ \bibinfo {pages} {13} (\bibinfo {year}
  {1991})}\BibitemShut {NoStop}%
\bibitem [{\citenamefont {Angell}(1995)}]{angell1995}%
  \BibitemOpen
  \bibfield  {author} {\bibinfo {author} {\bibfnamefont {C.~A.}\ \bibnamefont
  {Angell}},\ }\href {\doibase 10.1126/science.267.5206.1924} {\bibfield
  {journal} {\bibinfo  {journal} {Science}\ }\textbf {\bibinfo {volume}
  {267}},\ \bibinfo {pages} {1924} (\bibinfo {year} {1995})}\BibitemShut
  {NoStop}%
\bibitem [{\citenamefont {G{\"{o}}tze}(2008)}]{gotzebook}%
  \BibitemOpen
  \bibfield  {author} {\bibinfo {author} {\bibfnamefont {W.}~\bibnamefont
  {G{\"{o}}tze}},\ }\href@noop {} {\emph {\bibinfo {title} {Complex Dynamics of
  Glass-Forming Liquids: A Mode-Coupling Theory}}}\ (\bibinfo  {publisher}
  {Oxford University Press},\ \bibinfo {year} {2008})\BibitemShut {NoStop}%
\bibitem [{\citenamefont {Das}(2004)}]{das2004}%
  \BibitemOpen
  \bibfield  {author} {\bibinfo {author} {\bibfnamefont {S.~P.}\ \bibnamefont
  {Das}},\ }\href {\doibase 10.1103/RevModPhys.76.785} {\bibfield  {journal}
  {\bibinfo  {journal} {Rev. Mod. Phys.}\ }\textbf {\bibinfo {volume} {76}},\
  \bibinfo {pages} {785} (\bibinfo {year} {2004})}\BibitemShut {NoStop}%
\bibitem [{\citenamefont {Gotze}\ and\ \citenamefont
  {Sjogren}(1992)}]{gotze1992}%
  \BibitemOpen
  \bibfield  {author} {\bibinfo {author} {\bibfnamefont {W.}~\bibnamefont
  {Gotze}}\ and\ \bibinfo {author} {\bibfnamefont {L.}~\bibnamefont
  {Sjogren}},\ }\href {\doibase 10.1088/0034-4885/55/3/001} {\bibfield
  {journal} {\bibinfo  {journal} {Rep. Prog. Phys.}\ }\textbf {\bibinfo
  {volume} {55}},\ \bibinfo {pages} {241} (\bibinfo {year} {1992})}\BibitemShut
  {NoStop}%
\bibitem [{\citenamefont {Reichman}\ and\ \citenamefont
  {Charbonneau}(2005)}]{reichman2005}%
  \BibitemOpen
  \bibfield  {author} {\bibinfo {author} {\bibfnamefont {D.~R.}\ \bibnamefont
  {Reichman}}\ and\ \bibinfo {author} {\bibfnamefont {P.}~\bibnamefont
  {Charbonneau}},\ }\href {\doibase 10.1088/1742-5468/2005/05/P05013}
  {\bibfield  {journal} {\bibinfo  {journal} {J. Stat. Mech.}\ ,\ \bibinfo
  {pages} {P05013}} (\bibinfo {year} {2005})}\BibitemShut {NoStop}%
\bibitem [{\citenamefont {Geszti}(1983)}]{geszti1983}%
  \BibitemOpen
  \bibfield  {author} {\bibinfo {author} {\bibfnamefont {T.}~\bibnamefont
  {Geszti}},\ }\href {\doibase 10.1088/0022-3719/16/30/010} {\bibfield
  {journal} {\bibinfo  {journal} {Journal of Physics C: Solid State Physics}\
  }\textbf {\bibinfo {volume} {16}},\ \bibinfo {pages} {5805} (\bibinfo {year}
  {1983})}\BibitemShut {NoStop}%
\bibitem [{\citenamefont {Pareek}\ \emph {et~al.}(2023)\citenamefont {Pareek},
  \citenamefont {Adhikari}, \citenamefont {Dasgupta},\ and\ \citenamefont
  {Nandi}}]{pareek2023}%
  \BibitemOpen
  \bibfield  {author} {\bibinfo {author} {\bibfnamefont {P.}~\bibnamefont
  {Pareek}}, \bibinfo {author} {\bibfnamefont {M.}~\bibnamefont {Adhikari}},
  \bibinfo {author} {\bibfnamefont {C.}~\bibnamefont {Dasgupta}}, \ and\
  \bibinfo {author} {\bibfnamefont {S.~K.}\ \bibnamefont {Nandi}},\ }\href
  {\doibase 10.1063/5.0166404} {\bibfield  {journal} {\bibinfo  {journal} {J.
  Chem. Phys.}\ }\textbf {\bibinfo {volume} {159}},\ \bibinfo {pages} {174503}
  (\bibinfo {year} {2023})}\BibitemShut {NoStop}%
\bibitem [{\citenamefont {Janssen}(2018)}]{janssen2018}%
  \BibitemOpen
  \bibfield  {author} {\bibinfo {author} {\bibfnamefont {L.~M.~C.}\
  \bibnamefont {Janssen}},\ }\href {\doibase 10.3389/fphy.2018.00097}
  {\bibfield  {journal} {\bibinfo  {journal} {Front. Phys.}\ }\textbf {\bibinfo
  {volume} {6}},\ \bibinfo {pages} {97} (\bibinfo {year} {2018})}\BibitemShut
  {NoStop}%
\bibitem [{\citenamefont {Kardar}(2019)}]{kardarbook}%
  \BibitemOpen
  \bibfield  {author} {\bibinfo {author} {\bibfnamefont {M.}~\bibnamefont
  {Kardar}},\ }\href@noop {} {\emph {\bibinfo {title} {Statistical Physics of
  Fields}}}\ (\bibinfo  {publisher} {Cambridge University Press},\ \bibinfo
  {year} {2019})\BibitemShut {NoStop}%
\bibitem [{\citenamefont {Stringer}\ \emph {et~al.}(2021)\citenamefont
  {Stringer}, \citenamefont {Wang}, \citenamefont {Michaelos},\ and\
  \citenamefont {Pachitariu}}]{stringer2021}%
  \BibitemOpen
  \bibfield  {author} {\bibinfo {author} {\bibfnamefont {C.}~\bibnamefont
  {Stringer}}, \bibinfo {author} {\bibfnamefont {T.}~\bibnamefont {Wang}},
  \bibinfo {author} {\bibfnamefont {M.}~\bibnamefont {Michaelos}}, \ and\
  \bibinfo {author} {\bibfnamefont {M.}~\bibnamefont {Pachitariu}},\ }\href
  {\doibase 10.1038/s41592-020-01018-x} {\bibfield  {journal} {\bibinfo
  {journal} {Nature Methods}\ }\textbf {\bibinfo {volume} {18}},\ \bibinfo
  {pages} {100} (\bibinfo {year} {2021})}\BibitemShut {NoStop}%
\bibitem [{\citenamefont {Giavazzi}\ \emph {et~al.}(2018)\citenamefont
  {Giavazzi}, \citenamefont {Malinverno}, \citenamefont {Scita},\ and\
  \citenamefont {Cerbino}}]{giavazzi2018}%
  \BibitemOpen
  \bibfield  {author} {\bibinfo {author} {\bibfnamefont {F.}~\bibnamefont
  {Giavazzi}}, \bibinfo {author} {\bibfnamefont {C.}~\bibnamefont
  {Malinverno}}, \bibinfo {author} {\bibfnamefont {G.}~\bibnamefont {Scita}}, \
  and\ \bibinfo {author} {\bibfnamefont {R.}~\bibnamefont {Cerbino}},\ }\href
  {\doibase 10.3389/fphy.2018.00120} {\bibfield  {journal} {\bibinfo  {journal}
  {Front. Phys.}\ }\textbf {\bibinfo {volume} {6}},\ \bibinfo {pages} {120}
  (\bibinfo {year} {2018})}\BibitemShut {NoStop}%
\bibitem [{\citenamefont {Ranft}\ \emph {et~al.}(2010)\citenamefont {Ranft},
  \citenamefont {Basan}, \citenamefont {Elgeti}, \citenamefont {Joanny},
  \citenamefont {Prost},\ and\ \citenamefont {J{\"{u}}licher}}]{ranft2010}%
  \BibitemOpen
  \bibfield  {author} {\bibinfo {author} {\bibfnamefont {J.}~\bibnamefont
  {Ranft}}, \bibinfo {author} {\bibfnamefont {M.}~\bibnamefont {Basan}},
  \bibinfo {author} {\bibfnamefont {J.}~\bibnamefont {Elgeti}}, \bibinfo
  {author} {\bibfnamefont {J.-F.}\ \bibnamefont {Joanny}}, \bibinfo {author}
  {\bibfnamefont {J.}~\bibnamefont {Prost}}, \ and\ \bibinfo {author}
  {\bibfnamefont {F.}~\bibnamefont {J{\"{u}}licher}},\ }\href {\doibase
  10.1073/pnas.1011086107} {\bibfield  {journal} {\bibinfo  {journal} {Proc.
  Nat. Acad. Sci.}\ }\textbf {\bibinfo {volume} {107}},\ \bibinfo {pages}
  {20863} (\bibinfo {year} {2010})}\BibitemShut {NoStop}%
\bibitem [{\citenamefont {Minc}\ \emph {et~al.}(2011)\citenamefont {Minc},
  \citenamefont {Burgess},\ and\ \citenamefont {Chang}}]{minc2011}%
  \BibitemOpen
  \bibfield  {author} {\bibinfo {author} {\bibfnamefont {N.}~\bibnamefont
  {Minc}}, \bibinfo {author} {\bibfnamefont {D.}~\bibnamefont {Burgess}}, \
  and\ \bibinfo {author} {\bibfnamefont {F.}~\bibnamefont {Chang}},\ }\href
  {\doibase 10.1016/j.cell.2011.01.016} {\bibfield  {journal} {\bibinfo
  {journal} {Cell}\ }\textbf {\bibinfo {volume} {144}},\ \bibinfo {pages} {414}
  (\bibinfo {year} {2011})}\BibitemShut {NoStop}%
\bibitem [{\citenamefont {Chen}\ \emph {et~al.}(1997)\citenamefont {Chen},
  \citenamefont {Mrksich}, \citenamefont {Huang}, \citenamefont {Whitesides},\
  and\ \citenamefont {Ingber}}]{chen1997}%
  \BibitemOpen
  \bibfield  {author} {\bibinfo {author} {\bibfnamefont {C.~S.}\ \bibnamefont
  {Chen}}, \bibinfo {author} {\bibfnamefont {M.}~\bibnamefont {Mrksich}},
  \bibinfo {author} {\bibfnamefont {S.}~\bibnamefont {Huang}}, \bibinfo
  {author} {\bibfnamefont {G.~M.}\ \bibnamefont {Whitesides}}, \ and\ \bibinfo
  {author} {\bibfnamefont {D.~E.}\ \bibnamefont {Ingber}},\ }\href {\doibase
  10.1126/science.276.5317.1425} {\bibfield  {journal} {\bibinfo  {journal}
  {Science}\ }\textbf {\bibinfo {volume} {276}},\ \bibinfo {pages} {1425}
  (\bibinfo {year} {1997})}\BibitemShut {NoStop}%
\bibitem [{\citenamefont {Jaiswal}\ \emph {et~al.}(2006)\citenamefont
  {Jaiswal}, \citenamefont {Agrawal},\ and\ \citenamefont
  {Sinha}}]{jaiswal2006}%
  \BibitemOpen
  \bibfield  {author} {\bibinfo {author} {\bibfnamefont {M.}~\bibnamefont
  {Jaiswal}}, \bibinfo {author} {\bibfnamefont {N.}~\bibnamefont {Agrawal}}, \
  and\ \bibinfo {author} {\bibfnamefont {P.}~\bibnamefont {Sinha}},\ }\href
  {\doibase 10.1242/dev.02243} {\bibfield  {journal} {\bibinfo  {journal}
  {Development}\ }\textbf {\bibinfo {volume} {133}},\ \bibinfo {pages} {925}
  (\bibinfo {year} {2006})}\BibitemShut {NoStop}%
\bibitem [{\citenamefont {Cates}\ and\ \citenamefont
  {Ramaswamy}(2006)}]{cates2006}%
  \BibitemOpen
  \bibfield  {author} {\bibinfo {author} {\bibfnamefont {M.~E.}\ \bibnamefont
  {Cates}}\ and\ \bibinfo {author} {\bibfnamefont {S.}~\bibnamefont
  {Ramaswamy}},\ }\href {\doibase 10.1103/PhysRevLett.96.135701} {\bibfield
  {journal} {\bibinfo  {journal} {Phys. Rev. Lett.}\ }\textbf {\bibinfo
  {volume} {96}},\ \bibinfo {pages} {135701} (\bibinfo {year}
  {2006})}\BibitemShut {NoStop}%
\bibitem [{\citenamefont {Ruscher}\ \emph {et~al.}(2021)\citenamefont
  {Ruscher}, \citenamefont {Ciarella}, \citenamefont {Luo}, \citenamefont
  {Janssen}, \citenamefont {Farago},\ and\ \citenamefont
  {Baschnagel}}]{ruscher2021}%
  \BibitemOpen
  \bibfield  {author} {\bibinfo {author} {\bibfnamefont {C.}~\bibnamefont
  {Ruscher}}, \bibinfo {author} {\bibfnamefont {S.}~\bibnamefont {Ciarella}},
  \bibinfo {author} {\bibfnamefont {C.}~\bibnamefont {Luo}}, \bibinfo {author}
  {\bibfnamefont {L.~M.~C.}\ \bibnamefont {Janssen}}, \bibinfo {author}
  {\bibfnamefont {J.}~\bibnamefont {Farago}}, \ and\ \bibinfo {author}
  {\bibfnamefont {J.}~\bibnamefont {Baschnagel}},\ }\href {\doibase
  10.1088/1361-648X/abc4cc} {\bibfield  {journal} {\bibinfo  {journal} {J.
  Phys.: Condens. Matter}\ }\textbf {\bibinfo {volume} {33}},\ \bibinfo {pages}
  {064001} (\bibinfo {year} {2021})}\BibitemShut {NoStop}%
\bibitem [{\citenamefont {Tong}\ \emph {et~al.}(2022)\citenamefont {Tong},
  \citenamefont {Singh}, \citenamefont {Sknepnek},\ and\ \citenamefont
  {Košmrlj}}]{tong2022}%
  \BibitemOpen
  \bibfield  {author} {\bibinfo {author} {\bibfnamefont {S.}~\bibnamefont
  {Tong}}, \bibinfo {author} {\bibfnamefont {N.~K.}\ \bibnamefont {Singh}},
  \bibinfo {author} {\bibfnamefont {R.}~\bibnamefont {Sknepnek}}, \ and\
  \bibinfo {author} {\bibfnamefont {A.}~\bibnamefont {Košmrlj}},\ }\href
  {\doibase 10.1371/journal.pcbi.1010135} {\bibfield  {journal} {\bibinfo
  {journal} {PLOS Computational Biology}\ }\textbf {\bibinfo {volume} {18}},\
  \bibinfo {pages} {1} (\bibinfo {year} {2022})}\BibitemShut {NoStop}%
\end{thebibliography}%


\begin{thebibliography}{26}%
	\makeatletter
	\providecommand \@ifxundefined [1]{%
		\@ifx{#1\undefined}
	}%
	\providecommand \@ifnum [1]{%
		\ifnum #1\expandafter \@firstoftwo
		\else \expandafter \@secondoftwo
		\fi
	}%
	\providecommand \@ifx [1]{%
		\ifx #1\expandafter \@firstoftwo
		\else \expandafter \@secondoftwo
		\fi
	}%
	\providecommand \natexlab [1]{#1}%
	\providecommand \enquote  [1]{``#1''}%
	\providecommand \bibnamefont  [1]{#1}%
	\providecommand \bibfnamefont [1]{#1}%
	\providecommand \citenamefont [1]{#1}%
	\providecommand \href@noop [0]{\@secondoftwo}%
	\providecommand \href [0]{\begingroup \@sanitize@url \@href}%
	\providecommand \@href[1]{\@@startlink{#1}\@@href}%
	\providecommand \@@href[1]{\endgroup#1\@@endlink}%
	\providecommand \@sanitize@url [0]{\catcode `\\12\catcode `\$12\catcode
		`\&12\catcode `\#12\catcode `\^12\catcode `\_12\catcode `\%12\relax}%
	\providecommand \@@startlink[1]{}%
	\providecommand \@@endlink[0]{}%
	\providecommand \url  [0]{\begingroup\@sanitize@url \@url }%
	\providecommand \@url [1]{\endgroup\@href {#1}{\urlprefix }}%
	\providecommand \urlprefix  [0]{URL }%
	\providecommand \Eprint [0]{\href }%
	\providecommand \doibase [0]{https://doi.org/}%
	\providecommand \selectlanguage [0]{\@gobble}%
	\providecommand \bibinfo  [0]{\@secondoftwo}%
	\providecommand \bibfield  [0]{\@secondoftwo}%
	\providecommand \translation [1]{[#1]}%
	\providecommand \BibitemOpen [0]{}%
	\providecommand \bibitemStop [0]{}%
	\providecommand \bibitemNoStop [0]{.\EOS\space}%
	\providecommand \EOS [0]{\spacefactor3000\relax}%
	\providecommand \BibitemShut  [1]{\csname bibitem#1\endcsname}%
	\let\auto@bib@innerbib\@empty
	\bibitem [{\citenamefont {Honda}\ and\ \citenamefont
		{Eguchi}(1980)}]{smhonda1980}%
	\BibitemOpen
	\bibfield  {author} {\bibinfo {author} {\bibfnamefont {H.}~\bibnamefont
			{Honda}}\ and\ \bibinfo {author} {\bibfnamefont {G.}~\bibnamefont {Eguchi}},\
	}\href {https://doi.org/10.1016/S0022-5193(80)80021-X} {\bibfield  {journal}
		{\bibinfo  {journal} {J. Theor. Biol.}\ }\textbf {\bibinfo {volume} {84}},\
		\bibinfo {pages} {575} (\bibinfo {year} {1980})}\BibitemShut {NoStop}%
	\bibitem [{\citenamefont {Marder}(1987)}]{smmarder1987}%
	\BibitemOpen
	\bibfield  {author} {\bibinfo {author} {\bibfnamefont {M.}~\bibnamefont
			{Marder}},\ }\href {https://doi.org/10.1103/PhysRevA.36.438} {\bibfield
		{journal} {\bibinfo  {journal} {Phys. Rev. A}\ }\textbf {\bibinfo {volume}
			{36}},\ \bibinfo {pages} {438(R)} (\bibinfo {year} {1987})}\BibitemShut
	{NoStop}%
	\bibitem [{\citenamefont {Graner}\ and\ \citenamefont
		{Glazier}(1992)}]{smgraner1992}%
	\BibitemOpen
	\bibfield  {author} {\bibinfo {author} {\bibfnamefont {F.}~\bibnamefont
			{Graner}}\ and\ \bibinfo {author} {\bibfnamefont {J.~A.}\ \bibnamefont
			{Glazier}},\ }\href {https://doi.org/10.1103/PhysRevLett.69.2013} {\bibfield
		{journal} {\bibinfo  {journal} {Phys. Rev. Lett.}\ }\textbf {\bibinfo
			{volume} {69}},\ \bibinfo {pages} {2013} (\bibinfo {year}
		{1992})}\BibitemShut {NoStop}%
	\bibitem [{\citenamefont {Glazier}\ and\ \citenamefont
		{Graner}(1993)}]{smglazier1993}%
	\BibitemOpen
	\bibfield  {author} {\bibinfo {author} {\bibfnamefont {J.~A.}\ \bibnamefont
			{Glazier}}\ and\ \bibinfo {author} {\bibfnamefont {F.}~\bibnamefont
			{Graner}},\ }\href {https://doi.org/10.1103/PhysRevE.47.2128} {\bibfield
		{journal} {\bibinfo  {journal} {Phys. Rev. E}\ }\textbf {\bibinfo {volume}
			{47}},\ \bibinfo {pages} {2128} (\bibinfo {year} {1993})}\BibitemShut
	{NoStop}%
	\bibitem [{\citenamefont {Hirashima}\ \emph {et~al.}(2017)\citenamefont
		{Hirashima}, \citenamefont {Rens},\ and\ \citenamefont
		{Merks}}]{smhirashima2017}%
	\BibitemOpen
	\bibfield  {author} {\bibinfo {author} {\bibfnamefont {T.}~\bibnamefont
			{Hirashima}}, \bibinfo {author} {\bibfnamefont {E.~G.}\ \bibnamefont
			{Rens}},\ and\ \bibinfo {author} {\bibfnamefont {R.~M.~H.}\ \bibnamefont
			{Merks}},\ }\href {https://doi.org/10.1111/dgd.12358} {\bibfield  {journal}
		{\bibinfo  {journal} {Develop. Growth Differ.}\ }\textbf {\bibinfo {volume}
			{59}},\ \bibinfo {pages} {329} (\bibinfo {year} {2017})}\BibitemShut
	{NoStop}%
	\bibitem [{\citenamefont {Hogeweg}(2000)}]{smhogeweg2000}%
	\BibitemOpen
	\bibfield  {author} {\bibinfo {author} {\bibfnamefont {P.}~\bibnamefont
			{Hogeweg}},\ }\href {https://doi.org/10.1006/jtbi.2000.1087} {\bibfield
		{journal} {\bibinfo  {journal} {J. Theor. Biol.}\ }\textbf {\bibinfo {volume}
			{203}},\ \bibinfo {pages} {317} (\bibinfo {year} {2000})}\BibitemShut
	{NoStop}%
	\bibitem [{\citenamefont {Farhadifar}\ \emph {et~al.}(2007)\citenamefont
		{Farhadifar}, \citenamefont {R{\"{o}}per}, \citenamefont {Aigouy},
		\citenamefont {Eaton},\ and\ \citenamefont
		{J{\"{u}}licher}}]{smfarhadifar2007}%
	\BibitemOpen
	\bibfield  {author} {\bibinfo {author} {\bibfnamefont {R.}~\bibnamefont
			{Farhadifar}}, \bibinfo {author} {\bibfnamefont {J.-C.}\ \bibnamefont
			{R{\"{o}}per}}, \bibinfo {author} {\bibfnamefont {B.}~\bibnamefont {Aigouy}},
		\bibinfo {author} {\bibfnamefont {S.}~\bibnamefont {Eaton}},\ and\ \bibinfo
		{author} {\bibfnamefont {F.}~\bibnamefont {J{\"{u}}licher}},\ }\href
	{https://doi.org/10.1016/j.cub.2007.11.049} {\bibfield  {journal} {\bibinfo
			{journal} {Curr. Biol.}\ }\textbf {\bibinfo {volume} {17}},\ \bibinfo {pages}
		{2095} (\bibinfo {year} {2007})}\BibitemShut {NoStop}%
	\bibitem [{\citenamefont {Staple}\ \emph {et~al.}(2010)\citenamefont {Staple},
		\citenamefont {Farhadifar}, \citenamefont {R{\"o}per}, \citenamefont
		{Aigouy}, \citenamefont {Eaton},\ and\ \citenamefont
		{J{\"u}licher}}]{smstaple2010}%
	\BibitemOpen
	\bibfield  {author} {\bibinfo {author} {\bibfnamefont {D.~B.}\ \bibnamefont
			{Staple}}, \bibinfo {author} {\bibfnamefont {R.}~\bibnamefont {Farhadifar}},
		\bibinfo {author} {\bibfnamefont {J.-C.}\ \bibnamefont {R{\"o}per}}, \bibinfo
		{author} {\bibfnamefont {B.}~\bibnamefont {Aigouy}}, \bibinfo {author}
		{\bibfnamefont {S.}~\bibnamefont {Eaton}},\ and\ \bibinfo {author}
		{\bibfnamefont {F.}~\bibnamefont {J{\"u}licher}},\ }\href
	{https://doi.org/10.1140/epje/i2010-10677-0} {\bibfield  {journal} {\bibinfo
			{journal} {The European Physical Journal E}\ }\textbf {\bibinfo {volume}
			{33}},\ \bibinfo {pages} {117} (\bibinfo {year} {2010})}\BibitemShut
	{NoStop}%
	\bibitem [{\citenamefont {Fletcher}\ \emph {et~al.}(2014)\citenamefont
		{Fletcher}, \citenamefont {Osterfield}, \citenamefont {Baker},\ and\
		\citenamefont {Shvartsman}}]{smfletcher2014}%
	\BibitemOpen
	\bibfield  {author} {\bibinfo {author} {\bibfnamefont {A.}~\bibnamefont
			{Fletcher}}, \bibinfo {author} {\bibfnamefont {M.}~\bibnamefont
			{Osterfield}}, \bibinfo {author} {\bibfnamefont {R.}~\bibnamefont {Baker}},\
		and\ \bibinfo {author} {\bibfnamefont {S.}~\bibnamefont {Shvartsman}},\
	}\href {https://doi.org/https://doi.org/10.1016/j.bpj.2013.11.4498}
	{\bibfield  {journal} {\bibinfo  {journal} {Biophys. J.}\ }\textbf {\bibinfo
			{volume} {106}},\ \bibinfo {pages} {2291} (\bibinfo {year}
		{2014})}\BibitemShut {NoStop}%
	\bibitem [{\citenamefont {Barton}\ \emph {et~al.}(2017)\citenamefont {Barton},
		\citenamefont {Henkes}, \citenamefont {Weijer},\ and\ \citenamefont
		{Sknepnek}}]{smbarton2017}%
	\BibitemOpen
	\bibfield  {author} {\bibinfo {author} {\bibfnamefont {D.~L.}\ \bibnamefont
			{Barton}}, \bibinfo {author} {\bibfnamefont {S.}~\bibnamefont {Henkes}},
		\bibinfo {author} {\bibfnamefont {C.~J.}\ \bibnamefont {Weijer}},\ and\
		\bibinfo {author} {\bibfnamefont {R.}~\bibnamefont {Sknepnek}},\ }\href
	{https://doi.org/10.1371/journal.pcbi.1005569} {\bibfield  {journal}
		{\bibinfo  {journal} {PLOS Computational Biology}\ }\textbf {\bibinfo
			{volume} {13}},\ \bibinfo {pages} {1} (\bibinfo {year} {2017})}\BibitemShut
	{NoStop}%
	\bibitem [{\citenamefont {Kasza}\ \emph {et~al.}(2014)\citenamefont {Kasza},
		\citenamefont {Farrell},\ and\ \citenamefont {Zallen}}]{smkasza2014}%
	\BibitemOpen
	\bibfield  {author} {\bibinfo {author} {\bibfnamefont {K.~E.}\ \bibnamefont
			{Kasza}}, \bibinfo {author} {\bibfnamefont {D.~L.}\ \bibnamefont {Farrell}},\
		and\ \bibinfo {author} {\bibfnamefont {J.~A.}\ \bibnamefont {Zallen}},\
	}\href {https://doi.org/10.1073/pnas.1400520111} {\bibfield  {journal}
		{\bibinfo  {journal} {Proc. Natl. Acad. Sci. (USA)}\ }\textbf {\bibinfo
			{volume} {111}},\ \bibinfo {pages} {11732} (\bibinfo {year}
		{2014})}\BibitemShut {NoStop}%
	\bibitem [{\citenamefont {Yan}\ and\ \citenamefont {Bi}(2019)}]{smyan2019}%
	\BibitemOpen
	\bibfield  {author} {\bibinfo {author} {\bibfnamefont {L.}~\bibnamefont
			{Yan}}\ and\ \bibinfo {author} {\bibfnamefont {D.}~\bibnamefont {Bi}},\
	}\href {https://doi.org/10.1103/PhysRevX.9.011029} {\bibfield  {journal}
		{\bibinfo  {journal} {Phys. Rev. X}\ }\textbf {\bibinfo {volume} {9}},\
		\bibinfo {pages} {011029} (\bibinfo {year} {2019})}\BibitemShut {NoStop}%
	\bibitem [{\citenamefont {Tong}\ \emph {et~al.}(2022)\citenamefont {Tong},
		\citenamefont {Singh}, \citenamefont {Sknepnek},\ and\ \citenamefont
		{Košmrlj}}]{smtong2022}%
	\BibitemOpen
	\bibfield  {author} {\bibinfo {author} {\bibfnamefont {S.}~\bibnamefont
			{Tong}}, \bibinfo {author} {\bibfnamefont {N.~K.}\ \bibnamefont {Singh}},
		\bibinfo {author} {\bibfnamefont {R.}~\bibnamefont {Sknepnek}},\ and\
		\bibinfo {author} {\bibfnamefont {A.}~\bibnamefont {Košmrlj}},\ }\href
	{https://doi.org/10.1371/journal.pcbi.1010135} {\bibfield  {journal}
		{\bibinfo  {journal} {PLOS Computational Biology}\ }\textbf {\bibinfo
			{volume} {18}},\ \bibinfo {pages} {1} (\bibinfo {year} {2022})}\BibitemShut
	{NoStop}%
	\bibitem [{\citenamefont {G{\"{o}}tze}(2008)}]{smgotzebook}%
	\BibitemOpen
	\bibfield  {author} {\bibinfo {author} {\bibfnamefont {W.}~\bibnamefont
			{G{\"{o}}tze}},\ }\href@noop {} {\emph {\bibinfo {title} {Complex Dynamics of
				Glass-Forming Liquids: A Mode-Coupling Theory}}}\ (\bibinfo  {publisher}
	{Oxford University Press},\ \bibinfo {year} {2008})\BibitemShut {NoStop}%
	\bibitem [{\citenamefont {Das}(2004)}]{smdas2004}%
	\BibitemOpen
	\bibfield  {author} {\bibinfo {author} {\bibfnamefont {S.~P.}\ \bibnamefont
			{Das}},\ }\href {https://doi.org/10.1103/RevModPhys.76.785} {\bibfield
		{journal} {\bibinfo  {journal} {Rev. Mod. Phys.}\ }\textbf {\bibinfo {volume}
			{76}},\ \bibinfo {pages} {785} (\bibinfo {year} {2004})}\BibitemShut
	{NoStop}%
	\bibitem [{\citenamefont {Reichman}\ and\ \citenamefont
		{Charbonneau}(2005)}]{smreichman2005}%
	\BibitemOpen
	\bibfield  {author} {\bibinfo {author} {\bibfnamefont {D.~R.}\ \bibnamefont
			{Reichman}}\ and\ \bibinfo {author} {\bibfnamefont {P.}~\bibnamefont
			{Charbonneau}},\ }\href {https://doi.org/10.1088/1742-5468/2005/05/P05013}
	{\bibfield  {journal} {\bibinfo  {journal} {J. Stat. Mech.}\ ,\ \bibinfo
			{pages} {P05013}} (\bibinfo {year} {2005})}\BibitemShut {NoStop}%
	\bibitem [{\citenamefont {Hansen}\ and\ \citenamefont
		{McDonald}(2013)}]{smhansenmcdonald}%
	\BibitemOpen
	\bibfield  {author} {\bibinfo {author} {\bibfnamefont {J.-P.}\ \bibnamefont
			{Hansen}}\ and\ \bibinfo {author} {\bibfnamefont {I.~R.}\ \bibnamefont
			{McDonald}},\ }\href@noop {} {\emph {\bibinfo {title} {Theory of Simple
				Liquids}}},\ \bibinfo {edition} {4th}\ ed.\ (\bibinfo  {publisher}
	{Elsevier},\ \bibinfo {year} {2013})\BibitemShut {NoStop}%
	\bibitem [{\citenamefont {Franosch}\ \emph {et~al.}(1997)\citenamefont
		{Franosch}, \citenamefont {Fuchs}, \citenamefont {G{\"{o}}tze}, \citenamefont
		{Mayr},\ and\ \citenamefont {Singh}}]{smfranosch1997}%
	\BibitemOpen
	\bibfield  {author} {\bibinfo {author} {\bibfnamefont {T.}~\bibnamefont
			{Franosch}}, \bibinfo {author} {\bibfnamefont {M.}~\bibnamefont {Fuchs}},
		\bibinfo {author} {\bibfnamefont {W.}~\bibnamefont {G{\"{o}}tze}}, \bibinfo
		{author} {\bibfnamefont {M.~R.}\ \bibnamefont {Mayr}},\ and\ \bibinfo
		{author} {\bibfnamefont {A.~P.}\ \bibnamefont {Singh}},\ }\href
	{https://doi.org/10.1103/PhysRevE.55.7153} {\bibfield  {journal} {\bibinfo
			{journal} {Phys. Rev. Lett.}\ }\textbf {\bibinfo {volume} {55}},\ \bibinfo
		{pages} {7153} (\bibinfo {year} {1997})}\BibitemShut {NoStop}%
	\bibitem [{\citenamefont {Lubchenko}\ and\ \citenamefont
		{Wolynes}(2007)}]{smlubchenko2007}%
	\BibitemOpen
	\bibfield  {author} {\bibinfo {author} {\bibfnamefont {V.}~\bibnamefont
			{Lubchenko}}\ and\ \bibinfo {author} {\bibfnamefont {P.~G.}\ \bibnamefont
			{Wolynes}},\ }\href
	{https://doi.org/10.1146/annurev.physchem.58.032806.104653} {\bibfield
		{journal} {\bibinfo  {journal} {Ann. Rev. Phys. Chem.}\ }\textbf {\bibinfo
			{volume} {58}},\ \bibinfo {pages} {235} (\bibinfo {year} {2007})}\BibitemShut
	{NoStop}%
	\bibitem [{\citenamefont {Biroli}\ and\ \citenamefont
		{Bouchaud}(2012)}]{smbiroli2012}%
	\BibitemOpen
	\bibfield  {author} {\bibinfo {author} {\bibfnamefont {G.}~\bibnamefont
			{Biroli}}\ and\ \bibinfo {author} {\bibfnamefont {J.~P.}\ \bibnamefont
			{Bouchaud}},\ }in\ \href {https://doi.org/10.1002/9781118202470.ch2} {\emph
		{\bibinfo {booktitle} {Structural Glasses and Supercooled Liquids: Theory,
				Experiment, and Applications}}},\ \bibinfo {editor} {edited by\ \bibinfo
		{editor} {\bibfnamefont {P.~G.}\ \bibnamefont {Wolynes}}\ and\ \bibinfo
		{editor} {\bibfnamefont {V.}~\bibnamefont {Lubchenko}}}\ (\bibinfo {year}
	{2012})\BibitemShut {NoStop}%
	\bibitem [{\citenamefont {Kirkpatrick}\ and\ \citenamefont
		{Thirumalai}(2015)}]{smkirkpatrick2015}%
	\BibitemOpen
	\bibfield  {author} {\bibinfo {author} {\bibfnamefont {T.~R.}\ \bibnamefont
			{Kirkpatrick}}\ and\ \bibinfo {author} {\bibfnamefont {D.}~\bibnamefont
			{Thirumalai}},\ }\href {https://doi.org/10.1103/RevModPhys.87.183} {\bibfield
		{journal} {\bibinfo  {journal} {Rev. Mod. Phys.}\ }\textbf {\bibinfo
			{volume} {87}},\ \bibinfo {pages} {183} (\bibinfo {year} {2015})}\BibitemShut
	{NoStop}%
	\bibitem [{\citenamefont {Sadhukhan}\ and\ \citenamefont
		{Nandi}(2021)}]{smsadhukhan2021a}%
	\BibitemOpen
	\bibfield  {author} {\bibinfo {author} {\bibfnamefont {S.}~\bibnamefont
			{Sadhukhan}}\ and\ \bibinfo {author} {\bibfnamefont {S.~K.}\ \bibnamefont
			{Nandi}},\ }\href {https://doi.org/10.1103/PhysRevE.103.062403} {\bibfield
		{journal} {\bibinfo  {journal} {Phys. Rev. E}\ }\textbf {\bibinfo {volume}
			{103}},\ \bibinfo {pages} {062403} (\bibinfo {year} {2021})}\BibitemShut
	{NoStop}%
	\bibitem [{\citenamefont {Stringer}\ \emph {et~al.}(2021)\citenamefont
		{Stringer}, \citenamefont {Wang}, \citenamefont {Michaelos},\ and\
		\citenamefont {Pachitariu}}]{smstringer2021}%
	\BibitemOpen
	\bibfield  {author} {\bibinfo {author} {\bibfnamefont {C.}~\bibnamefont
			{Stringer}}, \bibinfo {author} {\bibfnamefont {T.}~\bibnamefont {Wang}},
		\bibinfo {author} {\bibfnamefont {M.}~\bibnamefont {Michaelos}},\ and\
		\bibinfo {author} {\bibfnamefont {M.}~\bibnamefont {Pachitariu}},\ }\href
	{https://doi.org/10.1038/s41592-020-01018-x} {\bibfield  {journal} {\bibinfo
			{journal} {Nature Methods}\ }\textbf {\bibinfo {volume} {18}},\ \bibinfo
		{pages} {100} (\bibinfo {year} {2021})}\BibitemShut {NoStop}%
	\bibitem [{\citenamefont {Pachitariu}\ and\ \citenamefont
		{Stringer}(2022)}]{smpachitariu2022}%
	\BibitemOpen
	\bibfield  {author} {\bibinfo {author} {\bibfnamefont {M.}~\bibnamefont
			{Pachitariu}}\ and\ \bibinfo {author} {\bibfnamefont {C.}~\bibnamefont
			{Stringer}},\ }\href {https://doi.org/10.1038/s41592-022-01663-4} {\bibfield
		{journal} {\bibinfo  {journal} {Nature Methods}\ }\textbf {\bibinfo {volume}
			{19}},\ \bibinfo {pages} {1634} (\bibinfo {year} {2022})}\BibitemShut
	{NoStop}%
	\bibitem [{\citenamefont {Ershov}\ \emph {et~al.}(2022)\citenamefont {Ershov},
		\citenamefont {Phan}, \citenamefont {Pylv{\"a}n{\"a}inen}, \citenamefont
		{Rigaud}, \citenamefont {Le~Blanc}, \citenamefont {Charles-Orszag},
		\citenamefont {Conway}, \citenamefont {Laine}, \citenamefont {Roy},
		\citenamefont {Bonazzi}, \citenamefont {Dum{\'e}nil}, \citenamefont
		{Jacquemet},\ and\ \citenamefont {Tinevez}}]{smershov2022}%
	\BibitemOpen
	\bibfield  {author} {\bibinfo {author} {\bibfnamefont {D.}~\bibnamefont
			{Ershov}}, \bibinfo {author} {\bibfnamefont {M.-S.}\ \bibnamefont {Phan}},
		\bibinfo {author} {\bibfnamefont {J.~W.}\ \bibnamefont
			{Pylv{\"a}n{\"a}inen}}, \bibinfo {author} {\bibfnamefont {S.~U.}\
			\bibnamefont {Rigaud}}, \bibinfo {author} {\bibfnamefont {L.}~\bibnamefont
			{Le~Blanc}}, \bibinfo {author} {\bibfnamefont {A.}~\bibnamefont
			{Charles-Orszag}}, \bibinfo {author} {\bibfnamefont {J.~R.~W.}\ \bibnamefont
			{Conway}}, \bibinfo {author} {\bibfnamefont {R.~F.}\ \bibnamefont {Laine}},
		\bibinfo {author} {\bibfnamefont {N.~H.}\ \bibnamefont {Roy}}, \bibinfo
		{author} {\bibfnamefont {D.}~\bibnamefont {Bonazzi}}, \bibinfo {author}
		{\bibfnamefont {G.}~\bibnamefont {Dum{\'e}nil}}, \bibinfo {author}
		{\bibfnamefont {G.}~\bibnamefont {Jacquemet}},\ and\ \bibinfo {author}
		{\bibfnamefont {J.-Y.}\ \bibnamefont {Tinevez}},\ }\href
	{https://doi.org/10.1038/s41592-022-01507-1} {\bibfield  {journal} {\bibinfo
			{journal} {Nature Methods}\ }\textbf {\bibinfo {volume} {19}},\ \bibinfo
		{pages} {829} (\bibinfo {year} {2022})}\BibitemShut {NoStop}%
	\bibitem [{\citenamefont {Schindelin}\ \emph {et~al.}(2012)\citenamefont
		{Schindelin}, \citenamefont {Arganda-Carreras}, \citenamefont {Frise},
		\citenamefont {Kaynig}, \citenamefont {Longair}, \citenamefont {Pietzsch},
		\citenamefont {Preibisch}, \citenamefont {Rueden}, \citenamefont {Saalfeld},
		\citenamefont {Schmid}, \citenamefont {Tinevez}, \citenamefont {White},
		\citenamefont {Hartenstein}, \citenamefont {Eliceiri}, \citenamefont
		{Tomancak},\ and\ \citenamefont {Cardona}}]{smschindelin2012}%
	\BibitemOpen
	\bibfield  {author} {\bibinfo {author} {\bibfnamefont {J.}~\bibnamefont
			{Schindelin}}, \bibinfo {author} {\bibfnamefont {I.}~\bibnamefont
			{Arganda-Carreras}}, \bibinfo {author} {\bibfnamefont {E.}~\bibnamefont
			{Frise}}, \bibinfo {author} {\bibfnamefont {V.}~\bibnamefont {Kaynig}},
		\bibinfo {author} {\bibfnamefont {M.}~\bibnamefont {Longair}}, \bibinfo
		{author} {\bibfnamefont {T.}~\bibnamefont {Pietzsch}}, \bibinfo {author}
		{\bibfnamefont {S.}~\bibnamefont {Preibisch}}, \bibinfo {author}
		{\bibfnamefont {C.}~\bibnamefont {Rueden}}, \bibinfo {author} {\bibfnamefont
			{S.}~\bibnamefont {Saalfeld}}, \bibinfo {author} {\bibfnamefont
			{B.}~\bibnamefont {Schmid}}, \bibinfo {author} {\bibfnamefont {J.-Y.}\
			\bibnamefont {Tinevez}}, \bibinfo {author} {\bibfnamefont {D.~J.}\
			\bibnamefont {White}}, \bibinfo {author} {\bibfnamefont {V.}~\bibnamefont
			{Hartenstein}}, \bibinfo {author} {\bibfnamefont {K.}~\bibnamefont
			{Eliceiri}}, \bibinfo {author} {\bibfnamefont {P.}~\bibnamefont {Tomancak}},\
		and\ \bibinfo {author} {\bibfnamefont {A.}~\bibnamefont {Cardona}},\ }\href
	{https://doi.org/10.1038/nmeth.2019} {\bibfield  {journal} {\bibinfo
			{journal} {Nature Methods}\ }\textbf {\bibinfo {volume} {9}},\ \bibinfo
		{pages} {676} (\bibinfo {year} {2012})}\BibitemShut {NoStop}%
\end{thebibliography}

\widetext
\clearpage

\hypersetup{
	colorlinks,
	linkcolor={red!90!black},
	citecolor={black!10!blue},
	urlcolor={blue!80!black}
}

\renewcommand{\r}{\textcolor{red}}
\renewcommand{\d}{\mathrm{d}}

\setcounter{equation}{0}
\setcounter{figure}{0}
\setcounter{section}{0}
\setcounter{table}{0}
\setcounter{page}{1}
\makeatletter
\renewcommand{\theequation}{S\arabic{equation}}
\renewcommand{\thefigure}{S\arabic{figure}}
\renewcommand{\thesection}{S\Roman{section}}
\renewcommand{\thesubsection}{S\Roman{section}\Alph{subsection}}
\renewcommand{\bibnumfmt}[1]{[S#1]}
\renewcommand{\citenumfont}[1]{S#1}

\begin{center}
	\textbf{\Large{Supplementary Material: The structure-dynamics feedback mechanism governs the glassy dynamics in epithelial monolayers}}
	
	\author{Satyam Pandey}
	\affiliation{Tata Institute of Fundamental Research, Hyderabad - 500046, India}
	
	\author{Soumitra Kolya}
	\affiliation{Tata Institute of Fundamental Research, Hyderabad - 500046, India}
	
	\author{Padmashree Devendran}
	\affiliation{Tata Institute of Fundamental Research, Hyderabad - 500046, India}
	
	\author{Souvik Sadhukhan}
	\affiliation{Tata Institute of Fundamental Research, Hyderabad - 500046, India}
	
	\author{Tamal Das}
	\email{tdas@tifrh.res.in}
	\affiliation{Tata Institute of Fundamental Research, Hyderabad - 500046, India}
	
	\author{Saroj Kumar Nandi}
	\email{saroj@tifrh.res.in}
	\affiliation{Tata Institute of Fundamental Research, Hyderabad - 500046, India}
	\vspace{16pt}
	
	Satyam Pandey, Soumitra Kolya, Padmashree Devendran, Souvik Sadhukhan, Tamal Das$^{*}$, and Saroj Kumar Nandi$^{*}$
	
	\vspace{2pt}
	
	\textit{Tata Institute of Fundamental Research, Hyderabad - 500046, India}
	
		
		\vspace{10pt}
		\hspace{10pt}	In this supplementary material, we provide the simulation details of Vertex models, a brief overview of MCT, the details of the numerical solution, a description of the non-ergodicity transition, wavevector-dependence of the MCT solution, the random first-order transition theory fits of the simulation data, and additional details of the experimental system and their analyses.
	\maketitle
\end{center}	
\section{Vertex model of confluent tissues}

The computational models of confluent cell monolayers have an energy function, Eq. (1) in the main text, and they represent cells as polygons \cite{smhonda1980,smmarder1987,smgraner1992,smglazier1993,smhirashima2017,smhogeweg2000,smfarhadifar2007,smstaple2010,smfletcher2014,smbarton2017}. These models can be both lattice-based or continuum. In this work, we have used a continuum model known as the Vertex model. In this model, the degrees of freedom are the vertices. The vertices in a cell are connected with a straight line to obtain the cell volume and the perimeter. 

\begin{figure}[h!]
	\centering
	\includegraphics[width=0.7\textwidth]{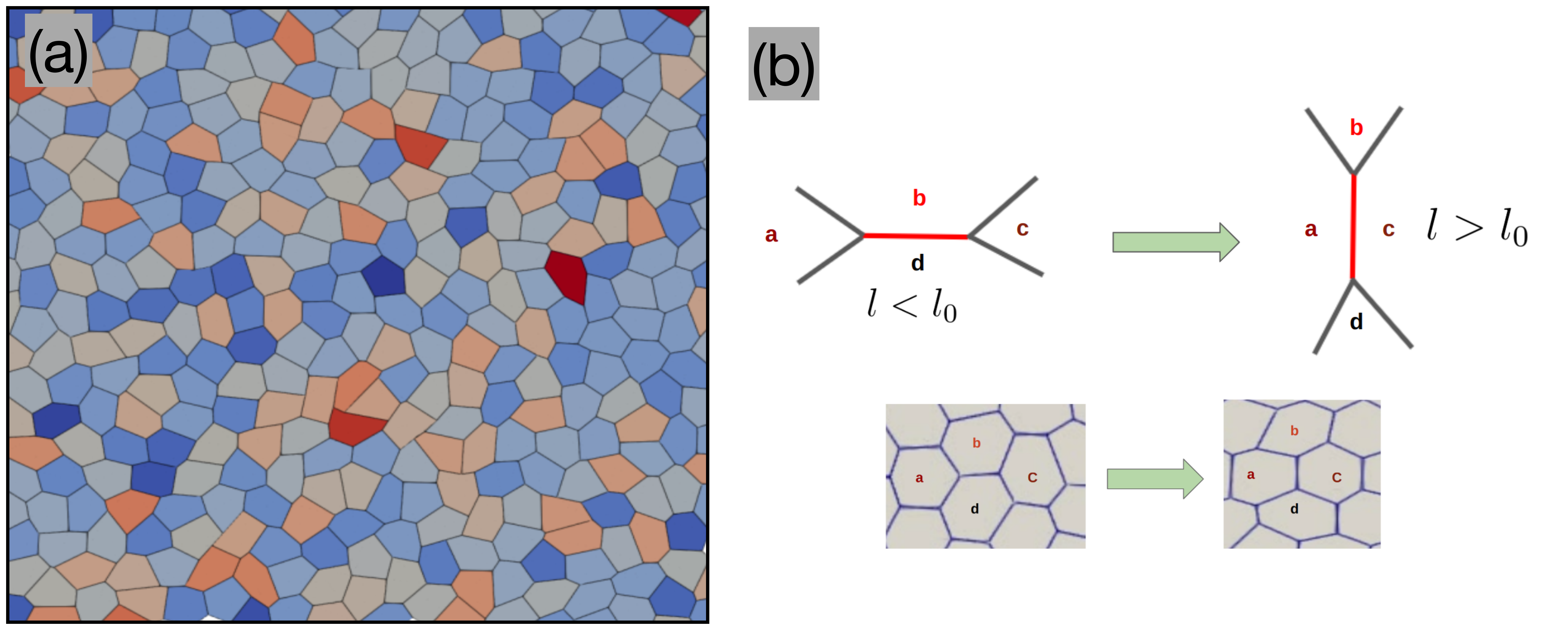}
	\caption{(a) A snapshot of the Vertex model from our simulation. The polygons represent different cells; colors are for visual presentation. (b) Schematic representation of the $T1$ transition or the neighbor exchange process.}
	\label{confluent}
\end{figure}

Figure \ref{confluent}(a) shows a typical configuration from our simulation. Unlike particulate systems, packing fraction in the confluent models remains one at all times and cannot be a control parameter.
One particular process, known as the $T1$ transition, is crucial for the dynamics in these systems. A $T1$ transition involves the disappearance of an edge between two neighboring cells followed by the subsequent formation of a perpendicular edge between them (Fig. \ref{confluent}b).

The Vertex model is one of the most widely used systems to study the glassy dynamics in confluent systems. In its usual implementation, three edges meet at each vertex (Fig. \ref{confluent}a). This property remains conserved throughout the dynamics. However, when $p_0>p_\text{min}$, vertices with more number of edges seem to become favorable \cite{smstaple2010,smkasza2014,smyan2019}. The consequence of this change in property is not yet entirely clear. Therefore, we restricted $p_0<p_\text{min}$ in our simulations.

\section{Simulation details}

We have used Brownian dynamics \cite{smtong2022} in our simulations:
\begin{equation}
	\gamma \dot{\mathbf{r}}_i=\mathbf{F}_i + \sqrt{2 D_T} \zeta,
\end{equation}
where $D_T$ is the translational diffusivity at temperature $T$, $\gamma$ is the substrate friction, and $\zeta$ is a random noise with zero mean and unit variance. The force, $\mathbf{F}_i$, is obtained via the energy function, $\mathcal{H}$, given in the main text \cite{smtong2022}.

We have used a square box of dimension $L=30$ for our simulations. We first generate a randomized point pattern comprising $N$ non-overlapping points via the random sequential addition algorithm. We use these points as the initial set of seed points for Voronoi tessellation incorporating periodic boundary conditions. We then equilibrate the configuration for the specific values of parameters and use it as the initial configuration for subsequent simulations. During the simulation, we monitored the edge lengths. If an edge becomes smaller than a specified length, $l_0$, we implemented a $T1$ transition. We ensured that the new edge length after the $T1$ transition was greater than $l_0$.

{\it Calculating the cell center:}
As described in the main text, we represent the cells by their centers (of mass). We compute the cell center from the vertex positions. Assuming the vertices come in either clockwise or counterclockwise order, we can calculate the position of the center as
\begin{equation}
	\begin{aligned} 
		& {\mathbf{r}^{\text{cm}}_{i,x}}=\frac{1}{6 A} \sum_{i=0}^{n-1}\left(x_i+x_{i+1}\right)\left(x_i y_{i+1}-x_{i+1} y_i\right) \\ 
		&  {\mathbf{r}^{\text{cm}}_{i,y}}=\frac{1}{6 A} \sum_{i=0}^{n-1}\left(y_i+y_{i+1}\right)\left(x_i y_{i+1}-x_{i+1} y_i\right),
	\end{aligned}
\end{equation}
where $n$ is the total number of vertices of a given cell, and the area $A$ is obtained as
\begin{equation}
	A=\frac{1}{2} \sum_{i=0}^{n-1}\left(x_i y_{i+1}-x_{i+1} y_i\right).
\end{equation}
As the cell perimeter is a closed loop object, we must have $x_0 = x_n$ and $y_0 = y_n$.

\section{Mode Coupling Theory (MCT)}
\subsection{Basic form of the theory}

The mode Coupling theory (MCT) was first derived for the glassy dynamics of particulate systems \cite{smgotzebook,smdas2004,smreichman2005}. It is a first-principle analytical theory for an immensely complex system. Here, we briefly highlight the main features of the theory. Consider the Hamiltonian $H$ for a particulate system,
\begin{equation}\label{particleH}
	H = \sum_i \frac{p_{i}^2}{2m} +\frac{1}{2}\sum_{i,j\neq i}\phi(r_{ij}),
\end{equation} 
where $p_i$ is the momentum of the $i$th particle (not to confuse with the perimeter of cell), $m$, the mass, $r_{ij}$, the inter-particle distance between the $i$th and the $j$th particles, $\phi$ is the interaction potential. We can then write down the equation of motion for any variable  $A(t)$ as 
\begin{equation}\label{liouville}
	\frac{dA(t)}{dt} = \{A(t),H\} = i\mathcal{L}A(t)
\end{equation}
where $\mathcal{L}$ is Liouville operator \cite{smreichman2005}. 
The number density in real space is
\begin{equation}
	\rho (r,t) = \sum_j^N \delta(r-r_{j}(t))
\end{equation}
and in Fourier space, $\rho(k,t)$ is
\begin{equation}
	\rho (k,t) = \sum_j^N \exp (ikr_{j}(t))
\end{equation}
where $N$ is the total number of particles and $r_{j}(t)$ is the position of particle $j$ at time $t$.
The intermediate scattering function $F(k,t)$ is 
\begin{equation}
	F(k,t) = \frac{1}{N} \langle \rho(-k,0) \rho(k,t) \rangle
\end{equation}
where the bracket denotes the ensemble average. The static structure factor $S(k)$ is
\begin{equation}
	S(k) = \frac{1}{N} \langle \rho(-k,0) \rho(k,0) \rangle.
\end{equation}

Using Eq. (\ref{liouville}) with $A(t)=(\rho(k,t), j(k,t))$, we can use the Mori-Zwanzig projection formalism to write down the equation of motion for $F(k,t)$ as 
\begin{equation}\label{mctequation}
	\frac{\partial}{\partial t} F(k,t)  + \frac{D_{0}k^{2}}{S(k)} F(k,t) + \int_{0}^{t} dt' M(k,t-t') \frac{\partial }{\partial t} F(k,t') = 0 ,
\end{equation}
where $S(k)$ is static structure factor (center-of-mass) of cells in confluent epithelial monolayer, $D_{0}$ is equal to $k_B T/m$ and $M(k,t)$ is the Memory kernel can be written as 
\begin{equation}
	M(k,t) = \frac{\rho D_{0}}{2} \int \frac{d\boldsymbol{q}}{(2\pi)^2} V_{k}^{2}(\boldsymbol{q},\mathbf{k}-\mathbf{q}) F(|\boldsymbol{k}-\boldsymbol{q}|,t) F(\boldsymbol{q},t),
\end{equation}
where $\rho$ is number density and $V_{k} $ is the vertex function can be written as 
\begin{equation}
	V({\boldsymbol{q},\mathbf{k}-\mathbf{q}}) =   \hat{k}.\boldsymbol{q}c(q) + \hat{k}.{(\mathbf{k}-\mathbf{q})}c(|\mathbf{k}-\mathbf{q}|)
\end{equation}
where $c(q)$ is the direct correlation function. Equation (\ref{mctequation}) is a non-linear integro-differential equation that we can self-consistently solve with a static structure factor as an input.
Note the generic features of the mode-coupling theory:
\begin{itemize}
	\item Equation (\ref{mctequation}) is quite generic: it is applicable for any dimension and any system in the absence of external fields.
	\item The potential $\phi$ can have an arbitrary form. Therefore, the theory is also applicable to the confluent system. $\phi$ in Eq. (\ref{particleH}) corresponds to $\mathcal{H}$ in the main text.
	\item The information of the system enters via $\phi$ alone; this is encoded through $S(k)$ in Eq. (\ref{mctequation}). Solution of the MCT equation requires $S(k)$ as the input, it acts as the initial condition, $F(k,t=0)=S(k)$.
\end{itemize}

\subsection{MCT in two-dimension}
We are interested in spatial dimension two ($2d$) for the confluent epithelial monolayers. Equation \ref{mctequation} can be written in $2d$ as,
\begin{equation}
	\tau_{k} \frac{\partial}{\partial t} f(k,t) + f(k,t) + \int_{0}^{t} \d t' m(k,t-t') \frac{\partial}{\partial t} f(k,t') = 0,
\end{equation}
where $\tau_{k} = S(k)/D_{0}k^2$, $f(k,t)= F(k,t)/S(k)$, and 
\begin{equation}
	\begin{aligned}
		m(k,t) = \frac{1}{2\rho k^2} \int\frac{\d\boldsymbol{q}}{(2\pi)^2} V_{k}^2(\mathbf{q},\mathbf{k}-\mathbf{q}) S(k) 
		S(|\mathbf{q}-\mathbf{k}|) S(q) f(|\mathbf{q}-\mathbf{k}|,t) f(\boldsymbol{q},t).
	\end{aligned}
\end{equation}
For the convenience of discretization, we rewrite the kernel in symmetrized form as
\begin{equation}
	\begin{aligned}
		m(k,t) = \frac{1}{2\rho k^2} \int \frac{\d\boldsymbol{q}}{(2\pi)^2} V^2({k/2 + q,k/2 - q})S(k) S(|\mathbf{k}/2 + \mathbf{q}|) S(|\mathbf{k}/2 - \mathbf{q}|) f(|\mathbf{k}/2 - \mathbf{q}|,t) f(|\mathbf{k}/2 - \mathbf{q}|,t).
	\end{aligned}
\end{equation}
We now use the change of variables as
\begin{align}
	x=|\mathbf{k}/2 + \mathbf{q}| = \sqrt{(q_{x}+k/2)^2 + q_{y}^2} = \sqrt{ q^2 + kq\cos\theta + \frac{k^2}{4}} \nonumber\\
	y=|\mathbf{k}/2 - \mathbf{q}| = \sqrt{(q_{x} - k/2)^2 + q_{y}^2} = \sqrt{ q^2 -kq\cos\theta + \frac{k^2}{4}},
\end{align}
where $\theta$ is the angle between $k$ and $q$.
As a result, the memory kernel becomes
\begin{align}\label{memorykernel}
	m(k,t) = \frac{\rho}{8 \pi^2 k^4} \int_{0}^{k_{m}} \d x \int_{|k-x|}^{|x+k|} \d y xy S(k) S(x) S(y)\frac{[(k^2 + x^2 - y^2)c(x) + (k^2 - x^2 + y^2)c(y)]^2}{(((4x^2k^2)-((k^2 + x^2- y^2)^2)^\frac{1}{2}) } f(x,t) f(y,t)
\end{align}
where $k_{m}$ is the numerical cut off wave vector and $c$ and $S$ are related by the
Ornstein-Zernike \cite{smhansenmcdonald} equation, 
\begin{equation}
	c(k) = \frac{1}{\rho} \left(1 - \frac{1}{S(k)}\right).
\end{equation}

{\it Details of the parameters for the numerical solution:}
For the numerical solution, we follow the algorithm developed by Fuchs et al. \cite{smfranosch1997}. $F(k,t)$ decays very fast at short times and quite slow at long times. Therefore, to numerically resolve both regimes, we must use an adaptive step size to discretize $t$. Thus, we start with a small time step $h=10^{-8}$ and double the step size every $N_t=128$ time step. We have discretized $k$ via an equally spaced grid of $N_{k}=200$ points with a grid spacing $\Delta k=0.2$. So our wave vector grid become $(i_{k} \times \Delta k)$ where $i_{k}$ runs from $1$ to $N_{k}$. 
We write the memory kernel, Eq. (\ref{memorykernel}), using the discrete Riemann sums,
\begin{equation}
	\int_{0}^{k_m} dx \int_{|k-x|}^{x+k} dy .... \rightarrow \sum_{i_{x} = 1 }^{N_{k}} (\Delta k) \sum_{y = |k-x|}^{|k+x|} (\Delta k).
\end{equation}
There is a divergence in the two extreme limits; we computed the sum by avoiding these points with respect to our $k$-discretization.

\subsection{The non-ergodicity transition of MCT}
As we have discussed in the main text, MCT works surprisingly well in a specific regime of parameter space. At high $T$, $F(k,t)$ quickly decays to zero. As we decrease $T$, the time evolution of $F(k,t)$ becomes slower. It first decays to a plateau and then towards zero at long times. As we further decrease $T$, $F(k,t)$ never becomes zero and remains stuck at a finite value forever; this is the non-ergodicity transition of MCT. The reason behind this transition remains unclear. It is generally believed that MCT is a mean-field theory; the sharp MCT transition becomes a crossover for finite-dimensional systems. However, the glassy properties are still governed by this genuine phase transition.

\begin{figure}[h!]
	\includegraphics[width=10cm]{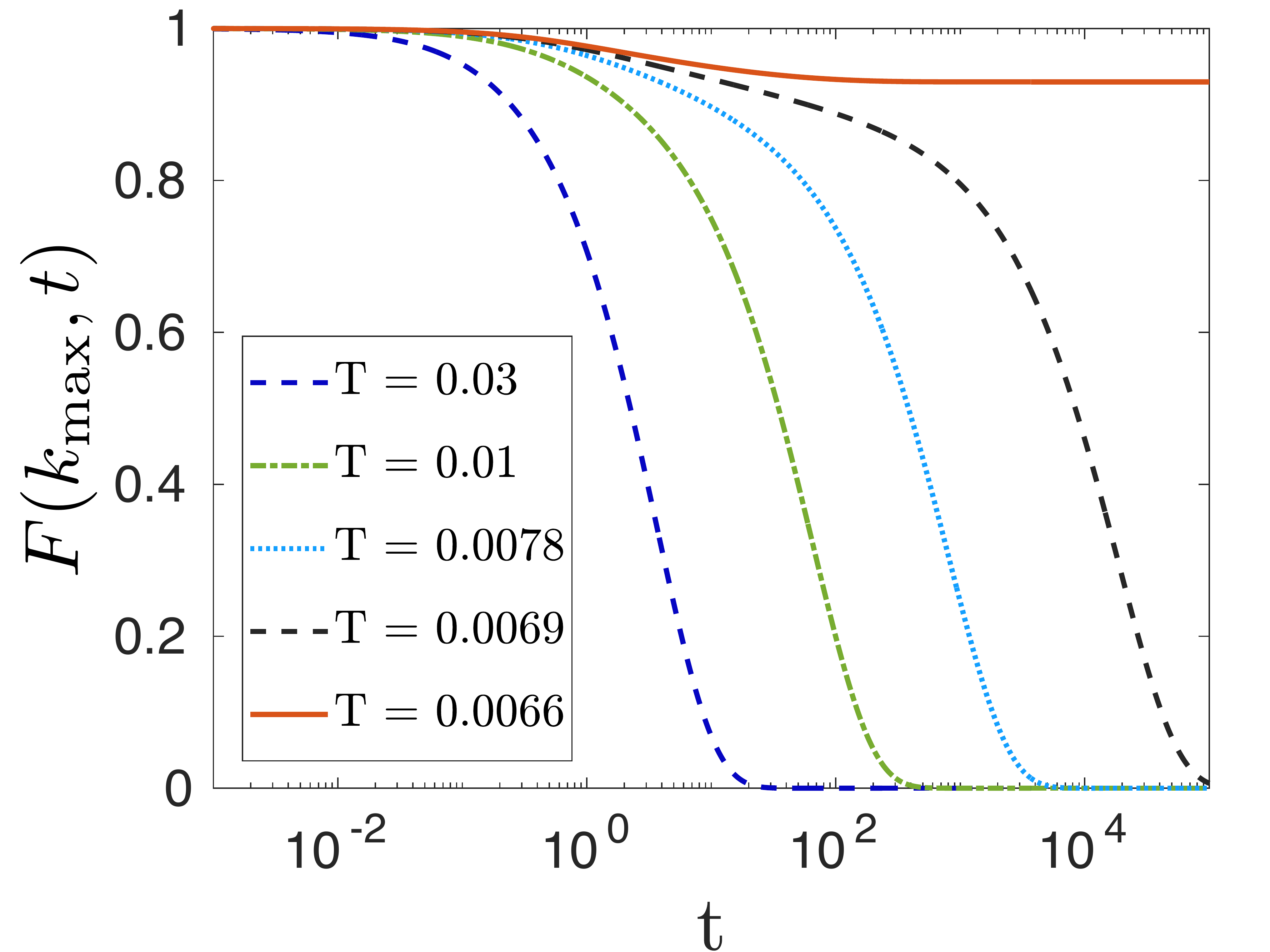}
	\caption{The non-ergodicity transition of MCT. As $T$ decreases, the decay of $F(k_\text{max},t)$ becomes slower. When $T=T_\text{\tiny MCT}$, the correlation function remains fixed at a finite value. }
	\label{nonergo}
\end{figure}

This transition should persist in confluent systems too. As Fig. \ref{nonergo} shows, the decay of $F(k_\text{max},t)$ becomes slower as $T$ becomes lower and eventually, at a low enough $T$ which is $T_\text{\tiny MCT}$, $F(k_\text{max},t)$  does not decay to zero.

\newpage
\subsection{$k$-dependence of $F(k,t)$}
We have discussed in the main text that the precise nature of the decay of $F(k,t)$ depends on the specific value of $k$. We have chosen six different values of $k$ and show the decay of $F(k,t)$ for these values of $k$.

\begin{figure*}[h!]
	\includegraphics[width=16cm]{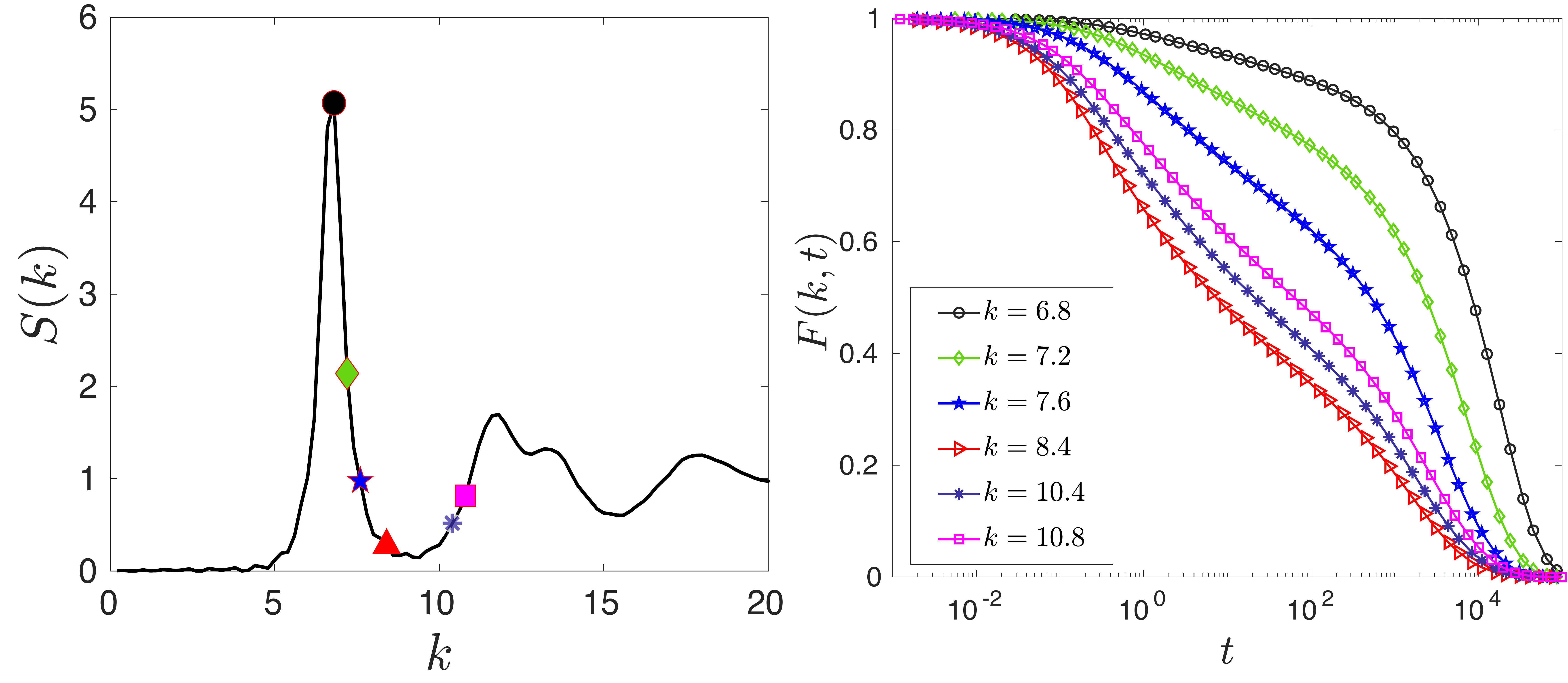}
	\caption{Wavevector-dependence of $F(k,t)$. (a) We use the $S(k)$ for $T=0.0069$ and $p_0=3.52$ as input to MCT to obtain $F(k,t)$. (b) The decay of $F(k,t)$ corresponding to the values of $k$ is shown by the symbols in (a). The precise nature of $F(k,t)$ depends on the specific values of $k$. }
	\label{kdependent}
\end{figure*}

\subsection{$k$ and $a$-dependence of the value of $\tau$ in MCT and simulation}

\begin{figure}[h!]
	\includegraphics[width=10cm]{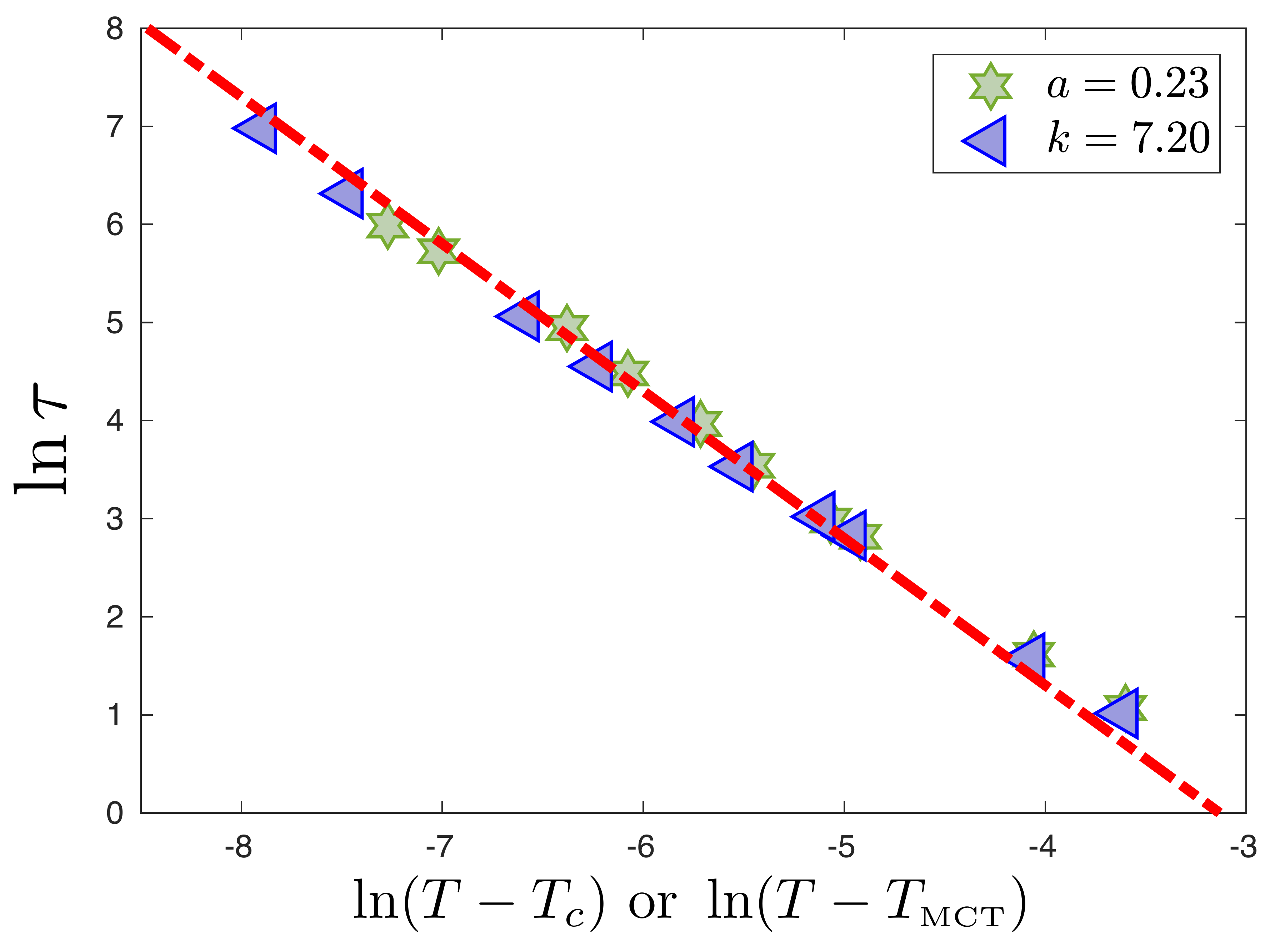}
	\caption{We have tuned $k$ and $a$ in such a way that the $\tau$ for the same parameters becomes the same. We have used $p_0=3.70$, $k=7.2$, and $a=0.23$.}
	\label{sametau}
\end{figure}

The precise values of $\tau$ in MCT and simulation depend on $k$ and $a$, respectively. However, their trends are independent of these parameters. Moreover, we can tune them $k$ and $a$ to have the same numerical values of $\tau$. The specific choices of $k$ and $a$ are motivated by practical considerations for better analysis. We demonstrate in Fig. \ref{sametau} that for suitable choices of $k$ and $a$, we can have similar values of $\tau$ within MCT and simulation. However, these parameters are not computationally convenient; therefore, we chose a different set of values, as stated in the main text.

\subsection{RFOT theory fits for the sub- and super-Arrhenius relaxations}
\begin{figure*}[h!]
	\includegraphics[width=14cm]{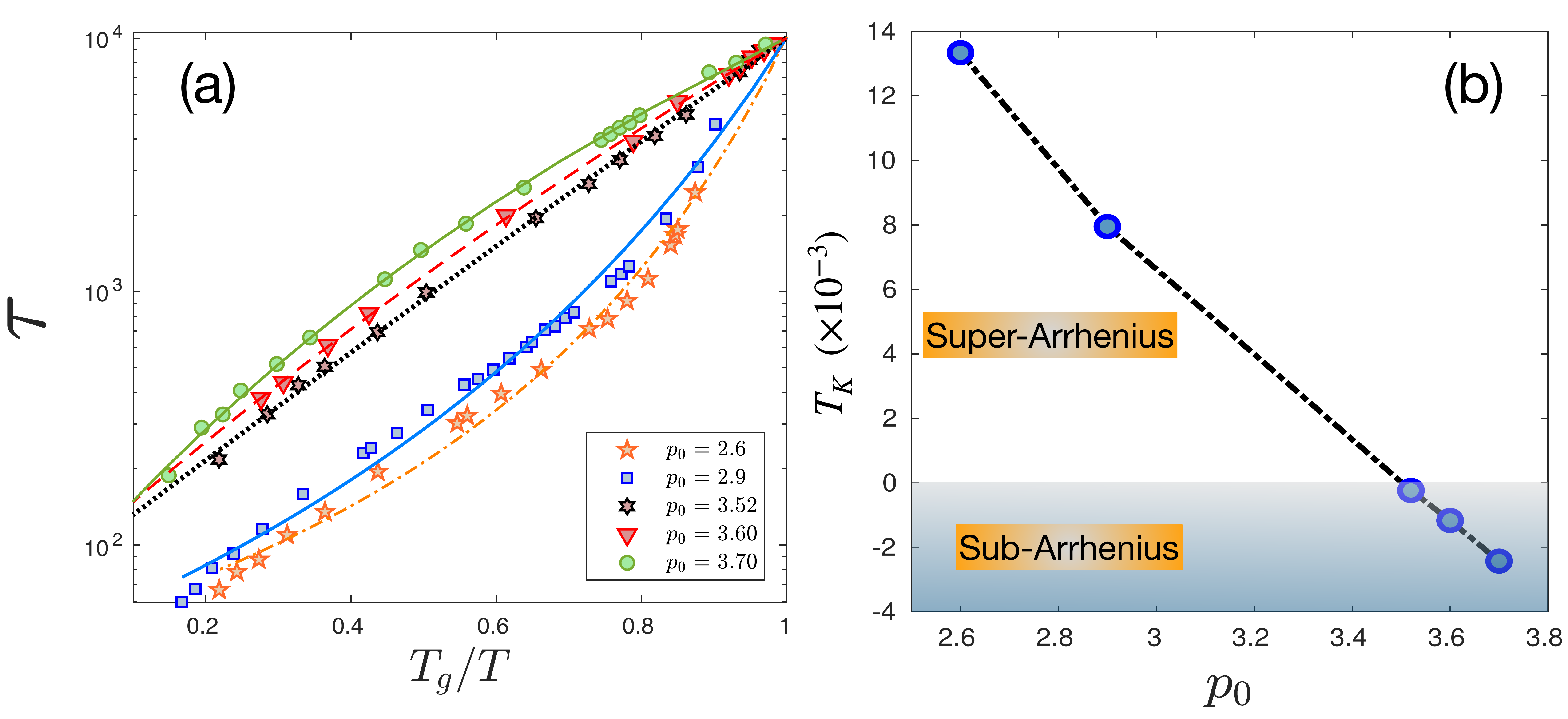}
	\caption{Application of the RFOT theory to the simulation data. (a) Fit of the RFOT theory form, Eq. (\ref{rfotform}), with the simulation data gives a reasonable description of the data. (b) $T_K$ in the sub-Arrhenius regime becomes negative, showing that the theory is not applicable in this regime.}
	\label{rfotfit}
\end{figure*}

Compared to the MCT, RFOT theory describes the relaxation dynamics as an activation process over a barrier. The relaxation time is obtained \cite{smlubchenko2007,smbiroli2012,smkirkpatrick2015} as
\begin{equation}\label{rfotform}
	\tau=\tau_0\exp\left[\frac{E}{T-T_K}\right],
\end{equation}
where $T_K$ is the Kauzmann temperature, $E$ is a constant, and $\tau_0$ is the high-$T$ value of $\tau$. RFOT theory crucially relies on the existence of a finite-temperature thermodynamic transition at $T_K$. As shown in Ref. \cite{smsadhukhan2021a}, we can phenomenologically extend the theory for confined systems.

Instead, we can consider an equilibrium scenario where confluency modifies the various constants. Then, we can fit Eq. (\ref{rfotform}) with the simulation data treating the constants $\tau_0$, $E$, and $T_K$ as fitting parameters. We show the fit in Fig. \ref{rfotfit}(a) and the values of $T_K$ in Fig. \ref{rfotfit}(b). Consistent with Ref. \cite{smsadhukhan2021a}, we find $T_K$ is negative in the sub-Arrhenius regime and positive in the super-Arrhenius regime, although the fits with the simulation data remain reasonable. The negative $T_K$ implies that the RFOT theory is not applicable in the sub-Arrhenius regime.

\section{Experimental details and additional results}

We conducted the cell culture experiment using Madin-Darby Canine Kidney (MDCK) epithelial cells in culture inserts with three wells (Ibidi chamber). We cultured the MDCK cells in Dulbecco's Modified Eagle Medium (DMEM) supplemented with 5\% fetal bovine serum (FBS) and 1\% antibiotic (penicillin and streptomycin). The incubation conditions are 37$^0 C$ and 5\% CO$_2$. We seeded the cells at different concentrations to achieve variable cell number densities. After the monolayer becomes confluent, we start the imaging process. Before imaging, we replaced the existing cell culture media with fresh media.

\subsection{Imaging}

We started imaging using a Leica DMi8 inverted microscope with a 20x objective lens. Images were captured in phase contrast mode at intervals of 2.5 minutes throughout a 4-hour duration. We show a fast-forwarded 4-hour-long imaging video in 6 seconds in Supplementary Movie (SI movie) I. Throughout the microscopy session, we maintained a stable environmental condition of $37^0C$ with $5\%$ CO$_2$, achieved by an incubation system mounted over the microscope. 

\subsection{Image Analysis and Cell Tracking}
After we have the microscopy images, we analyse them to obtain quantitative data. We conducted image analysis using the freely available Cellpose software \cite{smstringer2021} and an in-house custom-made code that we developed in MATLAB and Python. We primarily performed cellular segmentation using the Cellpose software and trained our custom segmentation model based on Cellpose 2.0 \cite{smpachitariu2022}. The performance of our trained model is robust; we show in Supplementary Movie II the track of the segmented images corresponding to the movie in SI Movie I. We tracked the cells using the TrackMate \cite{smershov2022} plugin in Fiji \cite{smschindelin2012}.

\begin{figure}
	\includegraphics[width=10cm]{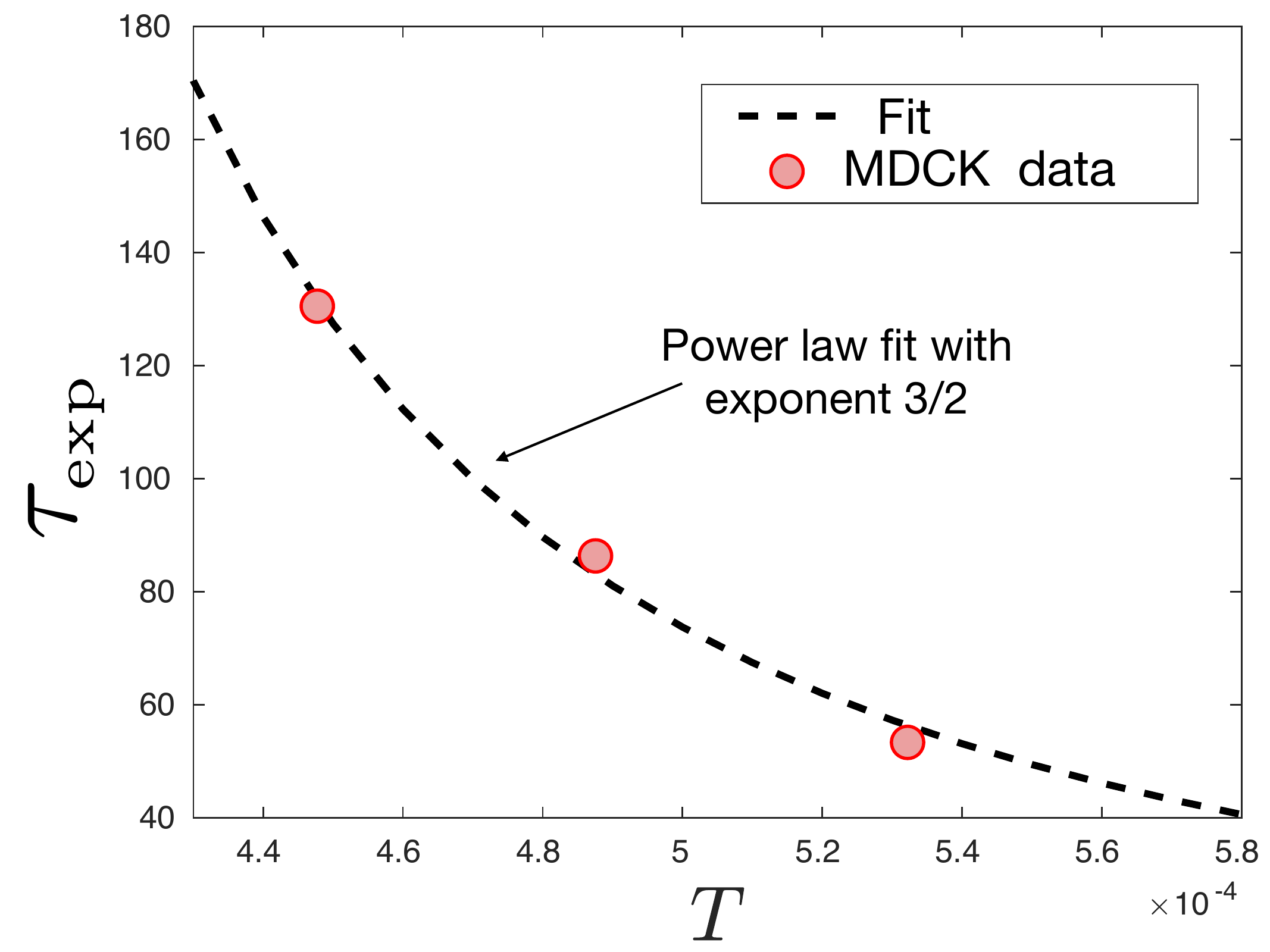}
	\caption{We fit the experimental data of relaxation time as a function of $T$ (obtained via the Cell Shape Theory, see main text for details) with a power law with exponent $3/2$ that MCT predicts. The data seems to be consistent with this prediction of the theory.}
	\label{powerlawexpt}
\end{figure}

\subsection{Power-law behavior of the relaxation time in experiment}
Since we could only get three data points for varying density, it is hard to reliably estimate the power law exponent from fit. However, we confirmed that the exponent 3/2, as we obtained from MCT, is consistent with the experimental data (Fig. \ref{powerlawexpt}).

\end{document}